# On nonperturbative localization with quasi-periodic potential

By J. Bourgain and M. Goldstein*

## I. Introduction, statements and preliminaries

Let $v$ be a real trigonometric polynomial on $\mathbb{T}^d$, the $d$-dimensional torus. Define for $\omega, \theta \in \mathbb{T}^d$,

$$A_n(\omega,\theta) = \begin{pmatrix} v(\theta+\omega) & 1 & 0 & 0 \\ 1 & v(\theta+2\omega) & 1 & \vdots \\ 0 & 1 & v(\theta+3\omega) & \vdots \\ 0 & 0 & 1 & \\ \vdots & \vdots & 0 & 1 \\ \vdots & \vdots & \vdots & v(\theta+n\omega) \end{pmatrix} \tag{0.1}$$

and the fundamental matrix

$$M_n(\omega,\theta,E) = \prod_{j=n}^{1} \begin{pmatrix} v(\theta+j\omega)-E & 1 \\ -1 & 0 \end{pmatrix}. \tag{0.2}$$

Then

$$M_n = \begin{bmatrix} \det(A_n(\omega,\theta)-E) & \det(A_{n-1}(\omega,\theta+\omega)-E) \\ -\det(A_{n-1}(\omega,\theta)-E) & -\det(A_{n-2}(\omega,\theta+\omega)-E) \end{bmatrix}. \tag{0.3}$$

Hence

$$\log\|M_n\| < Cn, \tag{0.4}$$

$$\log\|M_n^{-1}\| < Cn. \tag{0.5}$$

Define

$$L_n(\omega,E) = \frac{1}{n}\int_{\mathbb{T}^d} \log\|M_n(\omega,\theta,E)\| d\theta \tag{0.6}$$

*The first author was partially supported by NSF grant 9801013. The second author was supported in part by Ben-Gurion University and a grant of the Israel Academy of Sciences.

and

$$L(\omega, E) = \liminf_n \; L_n(\omega, E) \geq 0 \tag{0.7}$$

(the Lyapounov-exponent).

Our first result is the following:

THEOREM 1. *Let* $d = 1$ *or* $d = 2$ *and* $v$ *be as above. Assume that the Lyapounov exponent*

$$L(\omega, E) > 0 \;\; \textit{for all} \;\; \omega, E. \tag{0.8}$$

*Then, for almost all* $\omega$, *Anderson localization* (AL) *holds for the lattice Schrödinger operator* $A(\omega, 0)$, *where*

$$A_{jk}(\omega, 0) = v(j\omega)\delta_{jk} + (\delta_{j,k+1} + \delta_{j,k-1}), \quad j, k \in \mathbb{Z}. \tag{0.9}$$

Recall that Anderson localization means that $A(\omega, 0)$ has pure-point spectrum with exponentially decaying eigenfunctions.

It is well-known that to establish AL for $A$, it suffices to show that if $E \in \mathbb{R}$ and $\xi = (\xi_n)_{n \in \mathbb{Z}}$ at most of powerlike growth,

$$|\xi_n| \lesssim |n|^C \tag{0.10}$$

($\lesssim$ denotes inequality up to a multiplicative constant) satisfy the equation

$$(A - E)\xi = 0 \tag{0.11}$$

then $\xi$ decays exponentially

$$|\xi_n| \lesssim e^{-c|n|} \;\; \text{for some} \;\; c > 0. \tag{0.12}$$

*Comments.* (0) Theorem 1 for $d = 1$ was conjectured in [J].

(1) The statement of the theorem remains true for a real-analytic potential $v$. The argument given below does involve semi-algebraic sets. At this stage, some modifications are necessary, mainly involving suitable approximations by trigonometric polynomials. We also believe that the result remains valid in arbitrary dimension $d$ but did not carry out the details at this point.

(2) About condition (0.8). First, Theorem 1 remains valid if we assume $L(\omega, E) > 0$ for almost all $\omega$ (the 'almost all' independent of $E$).

If $v_0$ is an arbitrary nonconstant trigonometric polynomial (in any dimension), it follows from M. Herman's subharmonicity argument (see [H]) that (0.8) is satisfied for $v = \lambda v_0, \lambda > \lambda_0$. For $d = 1$, Sorets and Spencer [S-S] proved that more generally, for any nonconstant real analytic potential $v_0$ on $\mathbb{T}$, (0.8) holds for $v = \lambda v_0, \lambda > \lambda_0$ (this issue cannot be settled by straight forward approximation by trigonometric polynomials). The argument of [S-S] does not seem to extend easily to general real analytic potentials on $\mathbb{T}^d, d > 1$. We will give a different proof of this fact, based on the methods developed in this paper (in particular the large deviation theorem).

Thus, we have:

THEOREM 2. *If $v_0$ is a nonconstant real analytic potential on $\mathbb{T}^d$ and $\omega \in \mathbb{T}^d$ is a diophantine frequency vector, there is $\lambda_0$ such that $L(\omega, E) > \frac{1}{2}\log\lambda$ for $v = \lambda v_0, \lambda > \lambda_0$ and all $E$.*

(3) For $v_0 = \cos\theta$,

$$H_\lambda = \lambda\cos(n\omega + \theta) + \Delta \tag{0.13}$$

(the almost-Mathieu operator) satisfies (0.8) for $\lambda > 2$.

Very recently, S. Jitomirskaya [J] proved that for diophantine $\omega$ and almost all $\theta$, for $\lambda > 2$, AL holds. This is a particular case of our theorem, except for the fact that our result is in the measure category only (the considerations involved seem to require more than just diophantine assumptions).

(4) For $d = 2$, Theorems 1 and 2 (together with remark (2)) give the nonperturbative version of the Chulaevski-Sinai result ([C-S]).

(5) As will be clear below, our argument is based on a combination of measure information (large deviation estimates) and general facts on semi-algebraic sets (implying certain "complexity" bounds). It suggests a rather general scheme that one may try to apply in other related localization problems.

For instance, in [B-G-S], we use these methods to establish localization results for the skew shift.

It follows from (0.2) that

$$M_{n_1+n_2}(\omega, \theta, E) = M_{n_2}(\omega, \theta + n_1\omega, E)M_{n_1}(\omega, \theta, E); \tag{0.14}$$

hence

$$\log\|M_{n_1+n_2}(\omega, \theta, E)| \le \log\|M_{n_1}(\omega, \theta, E)\| + \log\|M_{n_2}(\omega, \theta + n_1\omega, E)\| \tag{0.15}$$

and by integration in $\theta$, it follows that

$$L_{n_1+n_2}(\omega, E) \le \frac{n_1}{n_1 + n_2}L_{n_1}(\omega, E) + \frac{n_2}{n_1 + n_2}L_{n_2}(\omega, E). \tag{0.16}$$

Thus, from (0.16)

$$L_n(\omega, E) \le L_m(\omega, E) \quad \text{if} \quad m < n, m|n \tag{0.17}$$

and

$$L_n(\omega, E) \le L_m(\omega, E) + C\frac{m}{n} \text{ if } m < n. \tag{0.18}$$

In particular, the sequence $\{L_n(\omega, E)\}$ converges to a limit $L(\omega, E) = \inf\limits_n L_n(\omega, E)$ for $n \to \infty$.

The interest of the products $M_n$ is to provide information about determinants, necessary to estimate Green's functions. Clearly

$$|\det(A_n(\omega, \theta) - E)| \le \|M_n(\omega, \theta, E)\| \lesssim |\det(A - E)| \tag{0.19}$$

for some

$$A \in \{A_n(\omega, \theta), A_{n-1}(\omega, \theta), A_{n-1}(\omega, \theta + \omega), A_{n-2}(\omega, \theta + \omega)\}. \tag{0.20}$$

Consider next the Green's function $\big(A_n(\omega, \theta) - E\big)^{-1}$.

From (0.1), it follows that if $1 \le n_1 \le n_2 \le n$, then the $(n_1, n_2)$-minor of $A_n(\omega, \theta) - E$ equals

$$\det[A_{n_1-1}(\omega, \theta) - E] \cdot \det[A_{n-n_2}(\omega, \theta + n_2\omega) - E]. \tag{0.21}$$

Hence by Cramer's rule

$$|[A_n(\omega, \theta) - E]^{-1}(n_1, n_2)| \le \frac{\|M_{n_1-1}(\omega, \theta, E)\| \cdot \|M_{n-n_2}(\omega, \theta + n_2\omega, E)\|}{|\det[A_n(\omega, \theta) - E]|}. \tag{0.22}$$

As we will show later, the upper-bound inequality (for large $m$)

$$\frac{1}{m} \log \|M_m(\omega, \theta, E)\| \le L_m(\omega, E) + o(1) \tag{0.23}$$

holds for arbitrary $\theta, E$, assuming $\omega$ diophantine. Assume $n \gg n_0 = n_0(\omega, E)$ where $n_0$ satisfies

$$L_{n_0}(\omega, E) < L(\omega, E) + o(1). \tag{0.24}$$

By (0.18), (0.19), (0.23), (0.24), the numerator of (0.22) may then be bounded by

$$e^{(n-|n_1-n_2|)L(\omega,E)+o(n)}. \tag{0.25}$$

Subject to replacement of $A_n(\omega, \theta)$ by one of the matrices in (0.20), we get thus

$$(0.22) < e^{n(L(\omega,E) - \frac{1}{n}\log\|M_n(\omega,\theta,E)\| + o(1))}\ e^{-|n_1-n_2|L(\omega,E)}. \tag{0.26}$$

Our main concern becomes therefore to obtain the lower bound

$$\frac{1}{n} \log \|M_n(\omega, \theta, E)\| > L_n(\omega, E) - o(1) \tag{0.27}$$

(which is a conditional issue with respect to parameters $\omega, \theta, E$).

If (0.27) holds, this will imply for some $A \in (0.20)$

$$|(A-E)^{-1}(n_1,n_2)| < e^{-L(\omega,E)|n_1-n_2|+o(n)}. \tag{0.28}$$

The relevance of the assumption

$$L(\omega,E) > 0 \tag{0.29}$$

becomes clear now, since it leads to the required exponential off-diagonal decay of the Green's function.

Denoting $A(\omega,\theta)$ the unrestricted matrix

$$A_{jk}(\omega,\theta) = v(\theta + j\omega)\delta_{jk} + \delta_{j,k-1} + \delta_{j,k+1} \quad (j,k \in \mathbb{Z}) \tag{0.30}$$

let, for $\Lambda \in \mathbb{Z}$,

$$A_\Lambda = R_\Lambda A R_\Lambda \qquad (R_\Lambda = \text{ restriction operator}) \tag{0.31}$$

and

$$G_\Lambda = (A_\Lambda - E)^{-1}. \tag{0.32}$$

Recall the resolvent identity

$$G_\Lambda = (G_{\Lambda_1} + G_{\Lambda_2}) - (G_{\Lambda_1} + G_{\Lambda_2})(A_\Lambda - A_{\Lambda_1} - A_{\Lambda_2})G_\Lambda \tag{0.33}$$

where $\Lambda \subset \mathbb{Z}$ is a disjoint union $\Lambda = \Lambda_1 \cup \Lambda_2$, provided the inverses make sense. One of the consequences of (0.33) is the following well-known "paving-property". Thus let $I \subset \mathbb{Z}$ be an interval of size $N > n$, such that for each $x \in \Lambda$, there is a size $n$-interval $I' \subset I$ satisfying

$$I' \supset \left\{ y \in I \mid |x-y| < \frac{n}{10} \right\}, \tag{0.34}$$

$$|G_{I'}(n_1,n_2)| < e^{-c|n_1-n_2|+o(n)}, \tag{0.35}$$

for some constant $c < 0$ and $n$ sufficiently large. Then also

$$|G_I(n_1,n_2)| < e^{-\frac{c}{2}|n_1-n_2|+o(n)}. \tag{0.36}$$

Full details on this argument will also appear in the last part (IV) of the paper. The remainder of the paper is divided into three parts containing the proof of Theorem 1 for $d = 1$ and $d = 2$ respectively and the proof of Theorem 2. In both cases $d = 1, 2$, the general scheme is the same but technically simpler for $d = 1$. The proof of Theorem 2 is given in the appendix.

We will denote in the sequel various constants by the same letter $C$.

## II. The case of 1-frequency

### 1. A large deviation estimate

The main ingredient providing estimates in measure is contained in the following:

LEMMA 1.1. *Assume* $\omega$ *satisfies a diophantine condition* (DC)

$$\|k\omega\| \gtrsim \frac{1}{|k|^{1+\varepsilon}} \text{ for } k \in \mathbb{Z}\backslash\{0\}. \tag{1.2}$$

*Then, for* $\sigma < \frac{1}{2}$ *and arbitrary* $E$, *with* $n$ *sufficiently large,*

$$\operatorname{mes}\left[\theta \in \mathbb{T} \left| \frac{1}{n} \log \|M_n(\omega, \theta, E)\| - L_n(\omega, E) \right| > n^{-\sigma}\right] < e^{-n^{1-2\sigma-}}. \tag{1.3}$$

*In general, if* $\omega$ *satisfies the condition* $\mathrm{DC}_{A,c}$

$$\|k\omega\| > c|k|^{-A} \quad \text{for} \quad k \in \mathbb{Z}\backslash\{0\} \tag{1.4}$$

*there is* $\sigma > 0$ *such that*

$$\operatorname{mes}\left[\theta \in \mathbb{T} \left| \frac{1}{\mathrm{n}} \log \|\mathrm{M_n}(\omega, \theta, \mathrm{E})\| - \mathrm{L_n}(\omega, \mathrm{E}) \right| > \mathrm{n}^{-\sigma}\right] < \mathrm{e}^{-\mathrm{n}^{\sigma}}. \tag{1.5}$$

*Proof.* By (0.2), $M_n(\omega, \cdot, E)$ clearly extends to an entire function $M_n(\omega, z, E)$ satisfying

$$\|M_n(\omega, z, E)\| + \|M_n^{-1}\| \lesssim (e^{C|\operatorname{Im} z|} + |E|)^n \tag{1.6}$$

(the constant $C$ in (1.6) depends on $v$; if, more generally, $v$ is real analytic, $z$ will be restricted to a strip $|\operatorname{Im} z| < \rho$, sufficient for our purpose).

Thus

$$\varphi(z) = \frac{1}{n} \log \|M_n(\omega, z, E)\| \tag{1.7}$$

is a subharmonic function, bounded on $D = [|z| < 1]$.

Let

$$G(z, w) = \log |z - w| - \log |1 - \bar{z}.w| \tag{1.8}$$

be the Green's function of $D$. Write for $z \in D$

$$\varphi(z) = \int_D \varphi(w) \Delta_w G(z, w) dw = \int_{\partial D} \varphi(s) \partial_n G(z, s) ds + \int_D G(z, w) \Delta\varphi(w) dw$$
$$= (1.9) + (1.10).$$

Note that (1.9) is the harmonic extension of a bounded function, hence smooth on $[|z| < \frac{3}{4}]$. In (1.10), $\Delta\varphi \geq 0$ by subharmonicity and $(\Delta\varphi)(w)dw$ defines a positive measure $d\mu$ of bounded mass on $D$.

The function $\varphi$ is 1-periodic on $\mathbb{R}$. We claim that

$$\widehat{\varphi}(k) = \int_{\mathbb{T}} \varphi(\theta)\, e^{-2\pi i k\theta} d\theta = O\Big(\frac{1}{|k|}\Big). \tag{1.11}$$

From the preceding and (1.8), it suffices clearly to verify that

$$\sup_{w\in D} \Big| \int_{\mathbb{R}} \log|x-w|\, e^{-2\pi i k x} \eta(x)dx \Big| \leq O\Big(\frac{1}{|k|}\Big) \tag{1.12}$$

where $\eta$ is a smooth bump-function satisfying

$$\operatorname{supp}\eta \subset \Big[-\frac{3}{4}, \frac{3}{4}\Big], \tag{1.13}$$

$$\sum_{s\in\mathbb{Z}} \eta(x+s) = 1, \text{ for all } x \in \mathbb{R}. \tag{1.14}$$

Thus (1.12) is equivalent to

$$\Big| \int \frac{x - \operatorname{Re} w}{|x-w|^2}\, e^{-2\pi i k x} \eta(x)dx \Big| \leq O(1), \tag{1.15}$$

proving (1.11).

Next, observe from (0.2) that for $r \geq 0$,

(1.16)

$$\begin{aligned} \|M_n(\omega, \theta + r\omega, E)\| \leq {} & \prod_{j=1}^{r} \Big\| \begin{pmatrix} v(\theta + j\omega) - E & 1 \\ -1 & 0 \end{pmatrix}^{-1} \Big\| \\ & \cdot\ \|M_n(\omega, \theta, E)\| \cdot \prod_{j=n+1}^{n+r} \Big\| \begin{pmatrix} v(\theta + j\omega) - E & 1 \\ -1 & 0 \end{pmatrix} \Big\| \\ \leq {} & C^r \|M_n(\omega, \theta, E)\|. \end{aligned}$$

Thus

$$|\varphi(\theta + r\omega) - \varphi(\theta)| \leq C\frac{|r|}{n}. \tag{1.17}$$

Expanding in Fourier series, we get

$$
\begin{aligned}
(1.20)\qquad |\varphi(\theta)-L_n| &\overset{(1.17)}{\leq} \left|\frac{1}{R}\sum_{r=1}^{R}\varphi(\theta+r\omega)-L_n\right|+O\Big(\frac{R}{n}\Big)\\
&\leq C\frac{R}{n}+\left|\sum_{|k|\geq 1}\widehat{\varphi}(k)\Big[\frac{1}{R}\sum_{1}^{R}e^{2\pi irk\omega}\Big]e^{2\pi ik\theta}\right|\\
&\overset{(1.11)}{\leq} C\frac{R}{n}+C\sum_{1\leq|k|\leq K}\frac{1}{|k|}\frac{1}{R\|k\omega\|+1}\\
&\qquad+\left|\sum_{|k|>K}\widehat{\varphi}(k)\Big[\frac{1}{R}\sum_{1}^{R}e^{2\pi irk\omega}\Big]e^{2\pi ik\theta}\right|\\
&= C\frac{R}{n}+(1.18)+(1.19)
\end{aligned}
$$

where the numbers $R<K$ are parameters to be specified.

Consider the sum (1.18). Write

$$
(1.21)\qquad (1.18)<C\sum_{\substack{\frac{1}{R}<\delta<1\\ \delta \text{ dyadic}}}\frac{1}{R\delta}\sum_{\substack{1\leq|k|\leq K\\ \|k\omega\|\sim\delta}}\frac{1}{|k|}.
$$

From the DC (1.2), it follows that if $\|k\omega\|<\delta$, then $|k|>\frac{1}{\delta^{1-}}$ and the second summation in (1.21) involves thus values of $k$ that are $\frac{1}{\delta^{1-}}$-separated. Hence

$$
(1.22)\qquad (1.18),(1.21)\leq C\sum_{\substack{\frac{1}{R}<\delta<1\\ \text{dyadic}}}\frac{1}{R\delta}\delta^{1-}\log K<\frac{\log K}{R^{1-}}.
$$

Also, by (1.11),

$$
(1.23)\qquad \int|(1.19)|d\theta\lesssim\Big(\sum_{|k|>K}\frac{1}{|k|^2}\Big)^{1/2}\lesssim K^{-1/2}.
$$

To prove the lemma, take thus

$$
(1.24)\qquad R=n^{1-\sigma},
$$

$$
(1.25)\qquad K=e^{n^{1-2\sigma-}}.
$$

Inequality (1.3) follows then from (1.20), (1.22), (1.23).

The proof of the inequality (1.5) under a weaker assumption (1.4) is similar.

*Remarks.* (1) It is clear from the previous argument that we only need the DC (1.2) or (1.4) with $k \in \mathbb{Z}$ restricted to $|k| < n$. This point will be of relevance later on.

(2) The method used here for $d = 1$ does not apply immediately in several variables and a general argument will be given in the discussion for the $d = 2$ case.

(3) More precise versions of Lemma 1.1 appear in the forthcoming paper [G-S].

## 2. The upper bound

We prove inequality (0.23).

LEMMA 2.1. *Assume $\omega$ satisfies* (1.4). *Then for all $\theta$ and $E$ in a bounded range*

$$\frac{1}{n} \log \|M_n(\omega, \theta, E)\| < L_n(\omega, E) + Cn^{-\sigma} \tag{2.2}$$

*for some $\sigma = \sigma(A) > 0$.*

*Proof.* Let $0 < \delta < 1$ be a small number to be specified. We majorize the function $\varphi(z) = \frac{1}{n} \log \|M_n(\omega, z, E)\|$ in (1.7) by the function $\varphi_1 \geq \varphi$ obtained by replacement in (1.8), (1.10) of the log function $\log |z - w|$ by $\log[|z - \omega| + \delta]$. This removes the singularity at 0.

Hence, we get now a Fourier transform estimate

$$\sup_{w \in D} \left| \int_{\mathbb{R}} \log[|x - w| + \delta] e^{-2\pi i k x} \eta(x) dx \right| < C \min\left(\frac{1}{|k|}, \frac{1}{\delta k^2}\right) \tag{2.3}$$

instead of (1.12).

Also

$$|\widehat{\varphi}_1(0) - \widehat{\varphi}(0)| \leq \sup_{w \in D} \int_{\mathbb{R}} \big| \log[|x - w| + \delta] - \log |x - w| \big| \eta(x) dx < \delta^{1-}. \tag{2.4}$$

Hence

$$\widehat{\varphi}_1(0) < L_n + \delta^{1-}. \tag{2.5}$$

Write

$$\begin{aligned}
\varphi(\theta) &\overset{(1.17)}{\leq} \frac{1}{R}\sum_1^R \varphi(\theta+r\omega) + O\Big(\frac{R}{n}\Big) \\
&\leq \frac{1}{R}\sum_1^R \varphi_1(\theta+r\omega) + O\Big(\frac{R}{n}\Big) \\
&\leq \widehat{\varphi}_1(0) + \sum_{k\neq 0} |\widehat{\varphi}_1(k)| \left|\frac{1}{R}\sum_1^R e^{2\pi i r k\omega}\right| + C\frac{R}{n} \\
&\overset{(2.5),(2.3)}{\leq} L_n + \delta^{1-} + C\sum_{1\leq |k|\leq K} \frac{1}{|k|}\frac{1}{R\|k\omega\|+1} + C\sum_{|k|>K}\frac{1}{k^2\delta} + \frac{CR}{n}
\end{aligned}$$

$$\overset{(1.4)}{<} L_n + \delta^{1-} + C\frac{\log K}{R^{1/A}} + \frac{C}{K\delta} + \frac{CR}{n}. \tag{2.6}$$

With $\delta = \frac{1}{n}, K = n^2, R = n^{1/2}$, (2.2) follows with $\sigma = \frac{1}{3A}$.

## 3. An averaging result

In order to settle a few issues later on, we will use the following property:

LEMMA 3.1. *Assume $\omega \in \mathbb{T}$ satisfies* $\mathrm{DC}_{A,c}$. *Then, for*

$$J > n^{2A}, \tag{3.2}$$

*there is the estimate*

$$\frac{1}{J}\sum_1^J \left(\frac{1}{n}\log\|M_n(\omega,\theta+j\omega,E)\|\right) = L_n(\omega,E) + O\Big(\frac{1}{n}\Big) \tag{3.3}$$

*uniformly for all $\theta$ and $E$ (in a bounded range).*

*Proof.* Clearly, from the definition of $M_n$,

$$\|\partial_\theta M_n(\omega,\theta,E)\| < C^n \tag{3.4}$$

and

$$\left|\partial_\theta\left[\frac{1}{n}\log\|M_n(\omega,\theta,E)\|\right]\right| < C^n \tag{3.5}$$

(where $C$ depends on $v$ and the range specified for $E$).

Hence, by (3.5), the function $\varphi$, specified in (1.7), satisfies

$$\sum_{k>K} |\widehat{\varphi}(k)| \lesssim K^{-1/2}\|\varphi'\|_2 < K^{-1/2}C^n. \tag{3.6}$$

Taking $K = C^{4n}$, we get for the left side of (3.3)

$$\frac{1}{J}\sum_{1}^{J}\varphi(\theta+j\omega) = L_n(\omega,E)+O\Big\{\sum_{1\le|k|\le K}|\widehat{\varphi}(k)|\frac{1}{1+J\|k\omega\|}+\sum_{|k|>K}|\widehat{\varphi}(k)|\Big\}$$

$$\overset{(3.6)}{=} L_n(\omega,E)+O\Big\{\sum_{1\le|k|\le K}\frac{1}{|k|(1+J\|k\omega\|)}+2^{-n}\Big\}$$

$$\overset{(1.4)}{=} L_n(\omega,E)+O\Big\{\frac{\log K}{J^{1/A}}+2^{-n}\Big\}. \tag{3.7}$$

With $J$ as in (3.2), (3.3) follows.

*Remark.* Observe again that we only use the more restricted assumption on $\omega$:

$$\|k\omega\| > c|k|^{-A} \quad \text{for} \quad 0<|k|<n^{2A}. \tag{3.8}$$

## 4. Elimination of the eigenvalue

LEMMA 4.1. *Let* $\log\log\bar{n} \ll \log n$. *Denote* $S\subset\mathbb{T}\times\mathbb{T}$ *the set of* $(\omega,\theta)$ *such that*

$$\|k\omega\| > c|k|^{-A} \text{ for } k\in\mathbb{Z},\ 0<|k|<n. \tag{4.2}$$

(4.3) *There is* $n_0<\bar{n}$ *and* $E$ *such that*

$$\|\big(A_{n_0}(\omega,0)-E\big)^{-1}\| > C^n \tag{4.4}$$

*and*

$$\frac{1}{n}\log\|M_n(\omega,\theta,E)\| < L_n(\omega,E)-n^{-\sigma}. \tag{4.5}$$

*Then*

$$\operatorname{mes} S < e^{-\frac{1}{2}n^{\sigma}} \tag{4.6}$$

(*for appropriate constants* $C,\sigma>0$, *in* (4.4)–(4.6)).

*Proof.* Denote

$$\Lambda_\omega = \bigcup_{n_0<\bar{n}} \text{Spec } A_{n_0}(\omega,0) \tag{4.7}$$

and

$$T_\omega = \Big\{\theta\in\mathbb{T}\Big|\max_{E_1\in\Lambda_\omega}\Big|\frac{1}{n}\log\|M_n(\omega,\theta,E_1)\|-L_n(\omega,E_1)\Big|>\frac{1}{2}n^{-\sigma}\Big\}. \tag{4.8}$$

Thus, by Lemma 1.1, assuming (4.2), taking into account the remark concerning condition (1.4), we have

$$\text{mes}\, T_\omega < \bar{n}^2 \bar{e}^{n^\sigma} < \bar{e}^{\frac{1}{2}n^\sigma}. \tag{4.9}$$

Hence (4.6) will follow from the fact that

$$S \subset \{(\omega, \theta) \in \mathbb{T} \times \mathbb{T} | \theta \in T_\omega\}. \tag{4.10}$$

Assume indeed that (4.4) and (4.5) hold for some $E$.

By (4.4), there is $E_1 \in \text{Spec}\, A_{n_0}(\omega, 0) \subset \Lambda_\omega$ such that

$$|E - E_1| < C^{-n}. \tag{4.11}$$

Clearly

$$\|M_n(\omega, \theta, E) - M_n(\omega, \theta, E_1)\| < C_v^n |E - E_1| < 1 \tag{4.12}$$

for appropriate $C$ in (4.4) and (4.11).

Hence

$$|\log \|M_n(\omega, \theta, E)\| - \log \|M_n(\omega, \theta, E_1)\| < 1 \tag{4.13}$$

and

$$|L_n(\omega, E) - L_n(\omega, E_1)| < \frac{1}{n}. \tag{4.14}$$

It follows from (4.5), (4.13), (4.14) that

$$\begin{aligned} \frac{1}{n} \log \|M_n(\omega, \theta, E_1)\| &< \frac{1}{n} \log \|M_n(\omega, \theta, E)\| + \frac{1}{n} \\ &< L_n(\omega, E) - n^{-\sigma} + \frac{1}{n} \\ &< L_n(\omega, E_1) - \frac{1}{2} n^{-\sigma}; \end{aligned} \tag{4.15}$$

hence $\theta \in T_\omega$.

This proves (4.10) and the lemma.

*Remark.* As will be clear later on, condition (4.4) will be fulfilled using the fact that the equation

$$\big(A(\omega, 0) - E\big)\xi = 0 \tag{4.16}$$

has a nontrivial solution, at most, of power-like growth $|\xi_k| \lesssim |k|^C$.

## 5. Semi-algebraic sets

In the next section, we will use the $(\omega, \theta)$-measure estimate obtained in Lemma 4.1 to get a statement only involving the frequency $\omega$. This step will be based on some additional considerations of the nature of conditions (4.4), (4.5). More specifically, we need to reformulate those conditions as polynomial inequalities in $\cos\omega, \sin\omega, \cos\theta, \sin\theta, E$ of not too high a degree.

When we denote the Hilbert-Schmidt norm of a matrix $B$ as

$$\|B\|_{\mathrm{HS}} = \Big(\sum_{i,j} |B_{ij}|^2\Big)^{1/2} \tag{5.1}$$

condition (4.4) may be replaced by

$$\sum_{1\le n_1,n_2\le n_0} \frac{\big(\det[(n_1,n_2) - \text{minor of } (A_{n_0}(\omega,0)-E)]\big)^2}{[\det(A_{n_0}(\omega,0)-E)]^2} \tag{5.2}$$
$$= \|[A_{n_0}(\omega,0)-E]^{-1}\|^2_{\mathrm{HS}} > C^{2n}.$$

Hence

$$\sum_{1\le n_1,n_2\le n_0} (\det[(n_1,n_2) - \text{minor of } \big(A_{n_0}(\omega,0)-E)]\big)^2 \tag{5.3}$$
$$> C^{2n}[\det(A_{n_0}(\omega,0)-E)]^2.$$

Condition (5.3) is of the form

$$P_1(\cos\omega, \sin\omega, E) > 0 \tag{5.4}$$

where $P_1(x_1,x_2,E)$ is a polynomial of degree at most $C_v n_0^2$ in $x_1, x_2$ and at most $2n_0$ in $E$; the constant $C_v$ depends on the trigonometric polynomial $v$ (in the general case of a real analytic potential $v$, one proceeds by truncation, replacing $v$ by a trigonometric polynomial $v_1$ of degree $< n^2$ say, with error $< e^{-cn^2}$, clearly sufficient in the context of conditions (4.4) or (4.5); this introduces an extra factor $n^2$ for the degree of $P_1$).

Consider next condition (4.5). In order to replace $L_n(\omega,E)$ using Lemma 3.1, we replace (4.2) by the stronger assumption (3.8); thus

$$\|k\omega\| > c|k|^{-A} \quad \text{for} \quad k\in\mathbb{Z}, 0<|k|<n^{2A}. \tag{5.5}$$

Since $J=[n^{2A}]$, it follows from (3.3) that condition (4.5) may be replaced by

$$\frac{1}{n}\log\|M_n(\omega,\theta,E)\| < \frac{1}{nJ}\sum_1^J \log\|M_n(\omega,j\omega,E)\| - n^{-\sigma}. \tag{5.6}$$

Hence

$$\|M_n(\omega,\theta,E)\|^{2J}_{\mathrm{HS}} < e^{-2Jn^{1-\sigma}} \prod_1^J \|M_n(\omega,j\omega,E)\|^2_{\mathrm{HS}}. \tag{5.7}$$

Thus (5.7) is of the form

$$P_2(\cos\omega,\sin\omega,\cos\theta,\sin\theta,E) > 0 \tag{5.8}$$

where $P_2(x_1, x_2, y_1, y_2, E)$ is a polynomial of degree at most $n^{C_A}$.

Thus conditions (4.4), with $n_0$ fixed, and (4.5) from Lemma 4.1 may be replaced by (5.4), (5.8), provided (4.2) is replaced by (5.5). If $\bar{n} < n^C$, these polynomials are of degree $< n^{C'}$.

Recall at this point the following general estimate on the number of connected components of semi-algebraic sets (Milnor, Thom).

PROPOSITION 5.9 (see [M]). *Let $V \subset \mathbb{R}^m$ be the set*

$$V = \bigcap [P_\alpha \overset{>}{(-)} 0] \tag{5.10}$$

*where the $P_\alpha$ are polynomials of degree $d_\alpha$. Let $d = \sum d_\alpha$. Then the number of components $\beta_0(V)$ of $V$, in particular, satisfies the inequality*

$$\beta_0(V) < d^m. \tag{5.11}$$

Assume the exponent $A$ in (5.5) is a fixed number. In the sequel, the letter $C$ may refer to different constants. For fixed $\theta$, apply Proposition 5.9 to the set

$$V = [P_1(x_1, x_2, E) > 0] \cap [P_2(x_1, x_2, \cos\theta, \sin\theta, E) > 0] \cap [x_1^2 + x_2^2 = 1] \tag{5.12}$$

which has thus at most $n^C$ components. Projecting out the variable $E$ (according to Lemma 4.1) we obtain at most $n^C$ intervals in $\mathbb{T}$. This property remains preserved if we take union over $n_0 < \bar{n} < n^C$ and also impose condition (5.5) on $\omega$. The conclusion is the following:

LEMMA 5.13. *In Lemma* 4.1, *take $\bar{n} < n^C$ and replace conditions* (4.4) *and* (4.5) *by the equivalent conditions* (5.4), (5.8). *Then, for fixed $\theta$, the set of $\omega$'s in* $[0, 1]$ *satisfying* (5.5) *and* (4.3) *is a union of at most $n^C$ intervals.*

This fact together with the measure estimate (4.6) are the main ingredients.

## 6. Frequency estimates

The information coming from Lemmas 4.1 and 5.13 will be combined with use of the following elementary fact.

LEMMA 6.1. *Let $S \subset \mathbb{T} \times \mathbb{T}$ be a set with the following property*:

(6.2) *For each $\theta \in \mathbb{T}$, the section $S_\theta = \{\omega \in \mathbb{T} | (\omega, \theta) \in S\}$ is a union of at most $M$ intervals.*

*Let $N \gg M$. Then,*

$$\text{mes}\,\{\omega \in [0,1] | (\omega, \ell\omega) \in S \ \ \textit{for some } \ell \sim N\} \leq N^3 (\text{mes}\, S)^{1/2} + MN^{-1}. \tag{6.3}$$

*Proof.* Estimate

$$\text{mes}\,\{\omega \in [0,1] |\, (\omega, \ell\omega) \in \text{S} \ \ \text{for some} \ \ \ell \sim \text{N}\} \tag{6.4}$$

$$\leq \sum_{\ell \sim N} \int_0^1 \chi_S(\omega, \ell\omega) d\omega \tag{6.5}$$

and by change of variable

$$\sum_{\ell \sim N} \sum_{0 \leq m < \ell} \frac{1}{\ell} \int_0^1 \chi_S \left( \frac{\theta + m}{\ell}, \theta \right) d\theta. \tag{6.6}$$

Fix $\theta$ and bound

$$\#\{(\ell, m) |\ |\ell|, |m| \lesssim N \ \ \text{and} \ \ \frac{\theta + m}{\ell} \in S_\theta\}. \tag{6.7}$$

Fix $0 < \gamma < \frac{1}{10N}$ and distinguish the following cases:

$$|S_\theta| > \gamma, \tag{6.8}$$

$$\inf_{0 < |k| \leq 2N} \|k\theta\| < 10 N^2 \gamma, \tag{6.9}$$

$$\text{negation of (6.8), (6.9).} \tag{6.10}$$

The contribution of (6.8), (6.9) to (6.6) is clearly bounded by

$$N\{\text{mes}\,[\theta \in [0,1] |\ |\text{S}_\theta| > \gamma] + \text{mes}\,[\theta \in [0,1] | \inf_{0<|\text{k}|<\text{N}} \|\text{k}\theta\| < 10\text{N}^2\gamma]\}$$

$$\leq CN(\gamma^{-1}|S| + N^3\gamma). \tag{6.11}$$

Assume (6.10). From assumption (6.2)

$$S_\theta = \bigcup_{\alpha=1}^{\beta} I_\alpha, \quad \beta \leq M, \tag{6.12}$$

where thus

$$|I_\alpha| \leq |S_\theta| \leq \gamma. \tag{6.13}$$

Fix $\alpha$ and suppose $(\ell_1, m_1) \neq (\ell_2, m_2)$ such that

$$\frac{\theta + m_1}{\ell_1} \in I_\alpha, \frac{\theta + m_2}{\ell_2} \in I_\alpha. \tag{6.14}$$

Then, by (6.13)

$$|(\ell_2 - \ell_1)\theta + (m_1\ell_2 - m_2\ell_1)| < N^2\gamma. \tag{6.15}$$

If $\ell_1 = \ell_2$, hence $m_1 \neq m_2$, it would follow that $1 \leq |m_1 - m_2| < 2N\gamma$ (impossible by the assumption on $\gamma$). If $\ell_1 \neq \ell_2$, (6.15) would imply that $\|(\ell_1 - \ell_2)\theta\| < N^2\gamma$ (impossible by negation of (6.9)). Consequently, if (6.10), each interval $I_\alpha$ in (6.12) contains at most one point $\frac{\theta+m}{\ell}$ and thus (6.7) $\leq M$.

From the preceding, it follows that

$$(6.6) \leq (6.11) + \frac{M}{N} \leq CN\gamma^{-1}|S| + N^4\gamma + N^{-1}M \tag{6.16}$$

and the lemma follows from appropriate choice of $\gamma$.

LEMMA 6.17. *Choose $\delta > 0$ and $n$ a sufficiently large integer. Denote $\Omega \subset \mathbb{T}$ the set of frequencies $\omega$ such that*

$$\omega \in DC_{10,c} \tag{6.18}$$

$$\text{There is } n_0 < n^C, \quad 2^{(\log n)^2} \leq \ell \leq 2^{(\log n)^3}, \text{ and } E \text{ such that} \tag{6.19}$$

$$\|[A_{n_0}(\omega, 0) - E]^{-1}\| > C^n, \tag{6.20}$$

$$\frac{1}{n}\log\|M_n(\omega, \ell\omega, E)\| < L_n(\omega, E) - \delta. \tag{6.21}$$

*Then*

$$\operatorname{mes}\Omega < 2^{-\frac{1}{4}(\log n)^2}. \tag{6.22}$$

*Proof.* Let $N$ be a range between $2^{(\log n)^2}$ and $2^{(\log n)^3}$ and let $\bar{n} = n^C$ in Lemma 4.1. The set $S \subset \mathbb{T} \times \mathbb{T}$ of frequencies $(\omega, \theta)$ satisfying (5.5) (with $A = 10$) and (5.4), (5.8) (hence (4.2), (4.4), (4.5)) is of measure

$$\operatorname{mes} \mathrm{S} < \mathrm{e}^{-\frac{1}{2}\mathrm{n}^\sigma} \tag{6.23}$$

for some $\sigma > 0$ (by (4.6)). Moreover, by Lemma 5.13, each of the sections $S_\theta$ of $S$ is a union of at most $n^C$ intervals. Hence Lemma 6.1 asserts that

(6.24)

$$\begin{aligned}
&\operatorname{mes}[\omega \in \mathbb{T} | \omega \text{ satisfies (6.18) and (6.19) for some } \mathrm{n}_0 < \mathrm{n}^{\mathrm{C}} \text{ and } \ell \sim \mathrm{N}] \\
&\qquad \leq \operatorname{mes}[\omega \in \mathbb{T} | (\omega, \ell\omega) \in \mathrm{S} \text{ for some } \ell \sim \mathrm{N}] \\
&\qquad \lesssim N^3 e^{-\frac{1}{4}n^\sigma} + n^C . N^{-1} \\
&\qquad < N^{-1/2}.
\end{aligned}$$

Summing over the ranges of $N$ implies the estimate (6.22).

## 7. Proof of the theorem in the 1-frequency case

Denote by $\Omega_{n,\delta}$ the frequency set obtained in Lemma 6.17. In fact, we replace (6.20) by the condition

$$\big\|\big(A_{[-n_0,n_0]}(\omega,0)-E\big)^{-1}\big\| > C^n \tag{7.1}$$

restricting the index set to $[-n_0,n_0]$ rather than to $[0,n_0]$.

Thus

$$\operatorname{mes}\Omega_{\mathrm{n},\delta} < \mathrm{e}^{-\frac{1}{4}(\log \mathrm{n})^2}. \tag{7.2}$$

Define

$$\Omega_\delta = \bigcap_{n'}\bigcup_{n>n'}\Omega_{n,\delta}, \quad \Omega=\bigcup_\delta \Omega_\delta \tag{7.3}$$

of measure

$$\operatorname{mes}\Omega_\delta \overset{(7.2)}{<} \inf_{\mathrm{n}'}\sum_{\mathrm{n}>\mathrm{n}'}\mathrm{e}^{-\frac{1}{4}(\log\mathrm{n})^2}=0, \quad \operatorname{mes}\Omega=0. \tag{7.4}$$

Assume $\omega\in \mathrm{DC}_{10,c}\backslash\Omega$ and let $E\in\mathbb{R}$, $\xi=(\xi_n)_{n\in\mathbb{Z}}$ satisfy the equation

$$\big(A(\omega,0)-E\big)\xi=0 \tag{7.5}$$

where

$$\xi_0=1 \text{ and } |\xi_n|\lesssim |n|^C. \tag{7.6}$$

Assume

$$L(\omega,E)>\delta_0>0. \tag{7.7}$$

Let $\delta=\frac{\delta_0}{1000}$. Since $\omega\notin\Omega_\delta$, there is $n'$ such that $\omega\notin\Omega_{n,\delta}$ for all $n>n'$. Assume moreover that

$$L_n(\omega,E)<L(\omega,E)+\delta \text{ for } n>n'. \tag{7.8}$$

Since (6.19) is not fulfilled, it follows that if for some $n_0<n^C$

$$\|G_{[-n_0,n_0]}\|>C^n \tag{7.9}$$

where $G_I=\big(A_I(\omega,0)-E\big)^{-1}$, then for all $2^{(\log n)^2}<|\ell|<2^{(\log n)^3}$,

$$\frac{1}{n}\log\|M_n(\omega,\ell\omega,E)\|>L_n(\omega,E)-\delta>\big(1-o(1)\big)L_n(\omega,E). \tag{7.10}$$

From the discussion (0.19)–(0.28) in the introduction, it follows that for each $2^{(\log n)^2}<\ell<2^{(\log n)^3}$, one of the matrices

$$A_n(\omega,\ell\omega), A_{n-1}(\omega,\ell\omega), A_{n-1}\big(\omega,(\ell+1)\omega\big), A_{n-2}\big(\omega,(\ell+1)\omega\big),$$

thus $A_I(\omega, 0)$ for some

$$I \in \{[\ell+1, \ell+n], [\ell+1, \ell+n-1], [\ell+2, \ell+n], [\ell+2, \ell+n-1]\}, \tag{7.11}$$

will satisfy

$$|G_I(n_1, n_2)| = |(A_I - E)^{-1}(n_1, n_2)| < e^{-L(\omega,E)|n_1-n_2|+o(n)} < e^{-\delta_0|n_1-n_2|+o(n)}. \tag{7.12}$$

Invoking also the paving property (0.33)–(0.36), one deduces from (7.12) that for $2^{(\log n)^2+1} < N < 2^{(\log n)^3-1}$ also the Green's function $G_{[\frac{N}{2}, 2N]}$ satisfies the estimate

$$|G_{[\frac{N}{2},2N]}(n_1, n_2)| < e^{-\delta|n_1-n_2|+n}. \tag{7.13}$$

Restricting the equation (7.5) to $[\frac{N}{2}, 2N]$ implies ($\xi_I \equiv R_I \xi$)

$$\big(A_{[\frac{N}{2},2N]}(\omega, 0) - E\big)\xi_{[\frac{N}{2},2N]} = -R_{[\frac{N}{2},2N]}\big(A(\omega, 0) - E\big)(\xi - \xi_{[\frac{N}{2},2N]}). \tag{7.14}$$

Hence, for $k \in [\frac{N}{2}, 2N]$

$$|\xi_k| \le \left|G_{[\frac{N}{2},2N]}\left(k, \frac{N}{2}\right)\right| \, |\xi_{\frac{N}{2}-1}| + |G_{[\frac{N}{2},2N]}(k, 2N)| \, |\xi_{2N+1}|, \tag{7.15}$$

$$|\xi_N| \overset{(7.6),(7.13)}{\le} N^C e^{n-\frac{\delta}{2}N} < e^{-\frac{\delta}{3}N}. \tag{7.16}$$

This is the required exponential decay property (also valid on the negative side).

It remains to show that (7.9) holds for some $n_0 < n^C$. Since

$$\begin{aligned} \xi_{[-n_0,n_0]} &= -G_{[-n_0,n_0]}R_{[-n_0,n_0]}\big(A(\omega, 0) - E\big)(\xi - \xi_{[-n_0,n_0]}), \\ 1 &\le \|G_{[-n_0,n_0]}\|(|\xi_{-n_0-1}| + |\xi_{n_0+1}|). \end{aligned} \tag{7.17}$$

Thus it will suffice to show that

$$\min_{|k|<n^C}(|\xi_k| + |\xi_{-k}|) < C^{-n}. \tag{7.18}$$

Let

$$n_1 = C\frac{n}{\delta}. \tag{7.19}$$

For $|k| < n^C$, it follows from the preceding that $|\xi_k| < C^{-n}$ provided that for some size $-n_1$ neighborhood $I$ of $k$, the Green's function $G_I$ will satisfy (7.12). Thus again from (0.27), (0.28), it clearly suffices to show that for some $0 < j < n^C$

$$\frac{1}{n_1}\log\|M_{n_1}(\omega, j\omega, E)\| = L_{n_1}(\omega, E) + o(1), \tag{7.20}$$

$$\frac{1}{n_1}\log\|M_{n_1}(\omega, (-j-n_1)\omega, E)\| = L_{n_1}(\omega, E) + o(1). \tag{7.21}$$

If (7.20) and (7.21) hold, then indeed

$$|\xi_k| + |\xi_{-k}| < n^C e^{-\delta_0 \frac{n_1}{2} + O(n_1)} < e^{-\frac{\delta_0}{3} n_1} < C^{-n} \text{ for } k = j + \left[\frac{n_1}{2}\right]. \tag{7.22}$$

Let $J = n^C$. We verify (7.20) and (7.21) by averaging over $j \in [J, 2J]$. Thus recalling Lemma 3.1 (with $n$ replaced by $n_1$), we see that

(7.23)

$$\frac{1}{J}\sum_{J+1}^{2J}\left[\frac{1}{n_1}\log\|M_{n_1}(\omega, j\omega, E)\| + \frac{1}{n_1}\log\|M_{n_1}\big(\omega, (-j-n_1)\omega, E\big)\|\right]$$
$$= 2L_{n_1}(\omega, E) + O\bigg(\frac{1}{n_1}\bigg).$$

Hence, there is $J < j \le 2J$ such that

$$\frac{1}{n_1}\log\|M_{n_1}(\omega, j\omega, E)\| + \frac{1}{n_1}\log\|M_{n_1}\big(\omega, (-j-n_1)\omega, E\big)\| \tag{7.24}$$
$$> 2L_{n_1}(\omega, E) + O\bigg(\frac{1}{n_1}\bigg),$$

implying by the upper bound (Lemma 2.1)

(7.25)

$$L_{n_1}(\omega, E) + o(n_1^{-\sigma}) > \frac{1}{n_1}\log\|M_{n_1}(\omega, j\omega, E)\|$$
$$> 2L_{n_1}(\omega, E) + O\bigg(\frac{1}{n_1}\bigg) - \big(L_{n_1}(\omega, E) + O(n_1^{-\sigma})\big)$$
$$= L_{n_1}(\omega, E) + O(n_1^{-\sigma}).$$

Similarly

$$\frac{1}{n_1}\log\|M_{n_1}\big(\omega, (-j-n_1)\omega, E\big)\| = L_{n_1}(\omega, E) + O(n_1^{-\sigma}). \tag{7.26}$$

This establishes (7.20) and (7.21). Hence (7.18) holds, which completes the argument.

It follows that Anderson localization (AL) holds for $A(\omega, 0)$ if $\omega \in DC_{10,c}\backslash\Omega$ and $L(\omega, E) > 0$.

## III. The 2-frequency case

Next, we carry out the preceding scheme for the Schrödinger operator $A(\omega, \theta) = v(\theta + j\omega)\delta_{jk} + \Delta$, where $\theta, \omega \in \mathbb{T}^2$. There is some extra work needed, both for the probabilistic and algebraic aspects. Let again $v = v(\theta_1, \theta_2)$ be a trigonometric polynomial on $\mathbb{T}^2$ and define $A_n(\omega, \theta), M_n(\omega, \theta, E), L_n(\omega, E)$, $L(\omega, E)$ as before. We first need the analogue of Lemma 1.1.

## 8. The large deviation estimate

LEMMA 8.1. *When* $\omega \in \mathbb{T}^2$ *satisfies a* $\mathrm{DC}_{A,c}$,

$$\|k.\omega\| > c|k|^{-A} \text{ for } k \in \mathbb{Z}^2\backslash\{0\}. \tag{8.2}$$

*Then*, *for some* $\sigma > 0$ *and* $n$ *sufficiently large*,

$$\operatorname{mes}\left[\theta \in \mathbb{T}^2 \,\middle|\, \left|\frac{1}{n}\log\|M_n(\omega,\theta,E)\| - L_n(\omega,E)\right| > n^{-\sigma}\right] < e^{-n^\sigma}. \tag{8.3}$$

*Proof.* Consider the function

$$\varphi(\theta_1,\theta_2) = \frac{1}{n}\log\|M_n(\omega,\theta,E)\|. \tag{8.4}$$

Then, for fixed $\theta_2$, $\varphi(\cdot,\theta_2)$ is the restriction to $\mathbb{R}$ of a subharmonic function and hence, by the argument in Lemma 1.1 verifies the Fourier coefficient estimates

$$\left\|\int_{\mathbb{T}}\varphi(\theta_1,\theta_2)e^{-2\pi ik\theta_1}d\theta_1\right\|_{L^\infty_{\theta_2}} \lesssim \frac{C}{|k|}. \tag{8.5}$$

Similarly

$$\left\|\int_{\mathbb{T}}\varphi(\theta_1,\theta_2)e^{-2\pi ik\theta_2}d\theta_2\right\|_{L^\infty_{\theta_1}} \le \frac{C}{|k|}. \tag{8.6}$$

Thus, if we consider a $K$-smoothing

$$\varphi_K(\theta) = \int_{\mathbb{T}^2}\varphi(\theta-\theta')P_K(\theta')d\theta' \tag{8.7}$$

it follows from (8.5), (8.6) that

$$\begin{aligned}\|\varphi-\varphi_K\|_2 &\sim \Bigg[\sum_{|k_1|+|k_2|\gtrsim K}|\widehat{\varphi}(k_1,k_2)|^2\Bigg]^{1/2} \\ &\le \Bigg[\sum_{|k_1|\gtrsim K}\left\|\int_{\mathbb{T}}\varphi(\theta_1,\cdot)e^{-2\pi ik_1\theta_1}\right\|^2_{L^2_{\theta_2}}\Bigg]^{1/2} \\ &\quad + \Bigg[\sum_{|k_2|\gtrsim K}\left\|\int_{\mathbb{T}}\varphi(\cdot,\theta_2)e^{-2\pi ik_2\theta_2}d\theta_2\right\|^2_{L^2_{\theta_1}}\Bigg]^{1/2} \\ &\le CK^{-1/2}.\end{aligned} \tag{8.8}$$

Recall also that

$$|\varphi(\theta+r\omega)-\varphi(\theta)| \le C\frac{|r|}{n}. \tag{8.9}$$

Hence

$$|\varphi_K(\theta + r\omega) - \varphi_K(\theta)| \le C\frac{|r|}{n}. \tag{8.10}$$

From (8.2), it follows that for any $\omega' \in \mathbb{T}^2$, there is $r \in \mathbb{Z}, |r| \lesssim \gamma^{-2-A}$ such that

$$\|r\omega - \omega'\| < \gamma. \tag{8.11}$$

Choose $\delta > 0$ and $r_1 \in \mathbb{Z}, 0 \le r_1 \lesssim \delta^{-2-A}$, such that

$$\|r_1\omega - (\delta, 0)\| < \frac{\delta}{10}. \tag{8.12}$$

Thus

$$r_1\omega \in \omega_1 + \mathbb{Z}^2,\ |\omega_1 - (\delta, 0)| < \frac{\delta}{10}. \tag{8.13}$$

From (8.10)

$$|\varphi_K(\theta + j\omega_1) - \varphi_K(\theta)| = |\varphi_K(\theta + r_1 j\omega) - \varphi_K(\theta)| \le C\frac{|jr_1|}{n}. \tag{8.14}$$

The function

$$\varphi_K(\theta + z\delta^{-1}\omega_1) = \int_{\mathbb{T}^2} \varphi(\theta - \theta' + z\delta^{-1}\omega_1) P_K(\theta') d\theta' \tag{8.15}$$

is a bounded subharmonic function for $z \in D(0, 2) \subset \mathbb{C}$. Hence, the restriction $\Phi$ of (8.15) to $[-1, 1]$, i.e.

$$\Phi(t) = \varphi_K(\theta + t\delta^{-1}\omega_1) \tag{8.16}$$

is essentially an average of functions

$$\log|t - w|,\ w \in D(0, 2).$$

Therefore

$$\|\Phi\|_{W^{s,1}[-1,1]} \lesssim (1 - s)^{-1} \quad \text{for} \quad s < 1. \tag{8.17}$$

Also, by (8.7)

$$\|\Phi\|_{C^2} \le \|\varphi_K\|_{C^2} \lesssim K^4. \tag{8.18}$$

Interpolation between (8.17) and (8.18) yields

$$\|\Phi\|_{W^{1,1}} \lesssim \log K. \tag{8.19}$$

Let $\zeta$ denote a smooth bump-function on $[-1, 1]$ such that for $\zeta \ge 0$,

$$\int_{-1}^{1} \zeta(t) dt = 1. \tag{8.20}$$

It follows from (8.19) that

$$\left|\int \Phi(t)\zeta(t)dt - \delta \sum_{|j|<\frac{1}{\delta}} \Phi(j\delta)\zeta(j\delta)\right| \lesssim \delta + \delta \int |\Phi'|\zeta \lesssim \delta \log K. \tag{8.21}$$

By (8.14), the expression

$$\delta \sum_{|j|<\frac{1}{\delta}} \Phi(j\delta)\zeta(j\delta) = \delta \sum_{|j|<\frac{1}{\delta}} \varphi_K(\theta + j\omega_1)\zeta(j\delta) = \varphi_K(\theta) + O\left(\frac{|r_1|}{\delta n} + \delta\right). \tag{8.22}$$

Hence

$$\varphi_K(\theta) = \int \varphi_K(\theta + t\delta^{-1}\omega_1)\zeta(t)dt + O\left(\frac{|r_1|}{\delta n} + \delta \log K\right). \tag{8.23}$$

Choosing next $r_2 \in \mathbb{Z}, |r_2| \lesssim \delta^{-2-A}$, we have

$$\|r_2\omega - (0,\delta)\| < \frac{\delta}{10}, \tag{8.24}$$

$$r_2\omega \in \omega_2 + \mathbb{Z}^2,\ |\omega_2 - (0,\delta)| < \frac{\delta}{10}. \tag{8.25}$$

Iteration of (8.23) gives

$$\varphi_K(\theta) = \iint \varphi_K(\theta + t_1\delta^{-1}\omega_1 + t_2\delta^{-1}\omega_2)\zeta(t_1)\zeta(t_2)dt_1dt_2 + O\left(\frac{\delta^{-3-A}}{n} + \delta \log K\right). \tag{8.26}$$

Observe that from (8.13), (8.25),

$$|\delta^{-1}\omega_1 - (1,0)| < \frac{1}{10}, \quad |\delta^{-1}\omega_2 - (0,1)| < \frac{1}{10}. \tag{8.27}$$

From (8.10) and (8.26), we obtain

$$\begin{aligned}\varphi_K(\theta) &= \frac{1}{r}\sum_1^r \varphi_K(\theta + j\omega) + O\left(\frac{r}{n}\right) \\ &= \frac{1}{r}\sum_1^r \iint \varphi_K(\theta + j\omega + t_1\delta^{-1}\omega_1 + t_2\delta^{-1}\omega_2)\zeta(t_1)\zeta(t_2)dt_1dt_2 \\ &\quad + O\left(\frac{r}{n} + \delta^{-3-A}n^{-1} + \delta \log K\right).\end{aligned} \tag{8.28}$$

By Fourier expansion, the first term of (8.28) equals

(8.29)

$$\widehat{\varphi}(0) + \sum_{k\in\mathbb{Z}^2\setminus\{0\}} \widehat{\varphi}_K(k)\frac{1}{r}\Big[\sum_1^r e^{2\pi ijk\omega}\Big]\widehat{\zeta}(k.\delta^{-1}\omega_1)\widehat{\zeta}(k.\delta^{-1}\omega_2)$$

$$= L_n(\omega, E) + O\Big(\sum_{k\in\mathbb{Z}^2\setminus\{0\}}(r\|k\omega\|+1)^{-1}\,|k|^{-C}\Big) \qquad \text{(by (8.27))}$$

$$= L_n(\omega, E) + O\Big(\frac{1}{r}\Big).$$

Thus

$$|\varphi_K(\theta) - L_n(\omega, E)| \lesssim \frac{1}{r} + \frac{r}{n} + \delta^{-3-A}n^{-1} + \delta\log K. \tag{8.30}$$

Appropriate choice of $r, \delta$ implies

$$\|\varphi_K - L_n(\omega, E)\|_\infty < (\log K).n^{-\frac{1}{4+A}}. \tag{8.31}$$

The lemma follows from (8.8), (8.31) by appropriate choice of $K$.

*Remark.* Previous arguments permit us to prove Lemma 8.1 in arbitrary dimension. The result again remains valid for real analytic $v$.

## 9. Averaging estimate

LEMMA 9.1. *Assume $\omega \in \mathbb{T}^2$ satisfies* (8.2). *Let*

$$J > n^{2A+5}. \tag{9.2}$$

*Then, for all $\theta \in \mathbb{T}^2$ and $E$,*

$$\frac{1}{nJ}\sum_1^J \log\|M_n(\omega, \theta + j\omega, E)\| = L_n(\omega, E) + O(n^{-1/2}). \tag{9.3}$$

*Proof.* Since $M_n(\omega, \theta, E)$ and $\varphi(\theta) = \frac{1}{n}\log\|M_n(\omega, \theta, E)\|$ have $\theta$-derivative bounded by $C^n$, cf. (3.4)–(3.6), we may identify $\varphi$ and $\varphi_K$ for $K = C^n$. Choose $\delta > 0$. With $\Phi$ as defined in (8.16), we have inequality (8.21)

$$\Big|\int \Phi(t)\zeta(t)dt - \delta\sum_{|j|<1/\delta}\Phi(j\delta)\zeta(j\delta)\Big| \lesssim \delta\log K \sim \delta n. \tag{9.4}$$

Hence, for all $\theta$,

$$\delta\sum_{|j|<\frac{1}{\delta}}\varphi(\theta + jr_1\omega)\zeta(j\delta) = \int\varphi_K(\theta + t\delta^{-1}\omega_1)\zeta(t)dt + O(\delta n) \tag{9.5}$$

where $r_1 \in \mathbb{Z}$, $|r_1| \leq \delta^{-2-A}$. Iteration gives again

$$\delta^2 \sum_{|j_1|,|j_2|\leq\frac{1}{\delta}} \varphi(\theta + (j_1 r_1 + j_2 r_2)\omega)\zeta(j_1\delta)\zeta(j_2\delta) \tag{9.6}$$

$$= \iint \varphi_K(\theta + t_1\delta^{-1}\omega_1 + t_2\delta^{-1}\omega_2)\zeta(t_1)\zeta(t_2)dt_1 dt_2 + O(\delta n)$$

$$\overset{(8.10)}{=} \frac{1}{r}\sum_1^r \iint \varphi_K(\theta + t_1\delta^{-1}\omega_1 + t_2\delta^{-1}\omega_2 + j\omega)\zeta(t_1)\zeta(t_2)dt_1 dt_2 + O\left(\frac{r}{n} + \delta n\right). \tag{9.7}$$

$$\overset{(8.29)}{=} L_n(\omega, E) + O\left(\frac{1}{r} + \frac{r}{n} + \delta n\right). \tag{9.8}$$

Choose

$$\delta = n^{-3/2},\ r = n^{1/2} \tag{9.9}$$

so that

$$(9.8) = L_n(\omega, E) + O(n^{-1/2}). \tag{9.10}$$

It follows that

$$\frac{1}{J}\sum_1^J \varphi(\theta + j\omega) \tag{9.11}$$

$$= \frac{\delta^2}{J}\sum_{j=1}^J \sum_{|j_1|,|j_2|\leq\frac{1}{\delta}} \varphi(\theta + (j + j_1 r_1 + j_2 r_2)\omega)\zeta(j_1\delta)\zeta(j_2\delta) + O\left(\delta + \frac{|r_1| + |r_2|}{\delta J}\right)$$

$$\overset{(9.8),(9.10)}{=} L_n(\omega, E) + O(n^{-1/2} + n^{\frac{3}{2}(3+A)}J^{-1}) \tag{9.12}$$

$$= L_n(\omega, E) + n^{-1/2} \tag{9.13}$$

by the choice (9.2) of $J$.

This proves the lemma.

## 10. Upper bound

We will use Lemma 9.1 to get the following substitute for Lemma 2.1.

LEMMA 10.1. *Assume $\omega \in \mathbb{T}^2$ satisfies* (8.2) *and let*

$$n > n_0^{2A+5}. \tag{10.2}$$

*Then*

$$\frac{1}{n}\log\|M_n(\omega,\theta,E)\| \le L_{n_0}(\omega,E) + O(n_0^{-1/2}). \tag{10.3}$$

*Proof.* In Lemma 9.1, replace $n$ by $n_0$ and $J$ by $n$. It follows from (9.3) that for some $0 \le r < n_0$,

$$\frac{1}{n}\sum_{q=0}^{n/n_0}\log\|M_{n_0}(\omega,\theta+(r+n_0q)\omega,E)\| \le L_{n_0}(\omega,E) + O(n_0^{-1/2}). \tag{10.4}$$

The left side of (10.4) is at least

$$\begin{aligned} &\frac{1}{n}\log\Bigg\|\prod_{q=[\frac{n}{n_0}]-1}^{0} M_{n_0}(\omega,\theta+(r+n_0q)\omega,E)\Bigg\| \\ &= \frac{1}{n}\log\Bigg\|\prod_{j=r+n_0[\frac{n}{n_0}]}^{r+1}\begin{pmatrix} v(\theta+j\omega)-E & -1\\ 1 & 0\end{pmatrix}\Bigg\| \\ &= \frac{1}{n}\log\|M_n(\omega,\theta,E)\| + O\Big(\frac{n_0}{n}\Big), \end{aligned} \tag{10.5}$$

implying (10.3).

*Remarks.* (1) Estimate (10.3) is weaker than (0.23) but clearly suffices to obtain estimate (0.25), provided we assume $n \gg n_0(\omega,E)^{2A+5}$, where $n_0(\omega,E)$ satisfies (0.24).

(2) Again for Lemmas 8.1, 9.2, 10.1 to hold, the assumption (8.2), i.e. $\omega \in DC_{A,c}$, may be replaced by the weaker hypothesis

$$\|k.\omega\| > c|k|^{-A} \text{ for } 0 < |k| < n^{C_A}. \tag{10.6}$$

## 11. Elimination of the eigenvalue

We have the analogue of Lemma 4.1.

LEMMA 11.1. *Let $\log\log\bar{n} \ll \log n$. Denote by $S \subset \mathbb{T}^2\times\mathbb{T}^2$ the set of $(\omega,\theta)$ such that*

$$\|k.\omega\| > c|k|^{-A} \textit{ for } k \in \mathbb{Z}^2, 0 < |k| < n^{C_A}. \tag{11.2}$$

(11.3) *Also there is* $n_0 < \bar{n}$ *and* $E$ *such that*

$$\|(A_{n_0}(\omega, 0) - E)^{-1}\| \geq C^n \tag{11.4}$$

*and*

$$\frac{1}{n} \log \|M_n(\omega, \theta, E)\| \leq L_n(\omega, E) - n^{-\sigma}. \tag{11.5}$$

*Then*

$$\operatorname{mes} \mathrm{S} < \mathrm{e}^{-\frac{1}{2}\mathrm{n}^{\sigma}} \tag{11.6}$$

(*for appropriate constants* $C, \sigma = \sigma(A) > 0$).

*Proof.* This is the same as for Lemma 4.1 when we replace Lemma 1.1 by Lemma 8.1.

## 12. Semi-algebraic sets

In Lemma 11.1, let the exponent $A$ be fixed and let $\bar{n} = n^C$ where $C$ is a sufficiently large constant (in fact, we use again the same letter $C$ to denote possibly different constants, provided there is no conflict). Let $J = n^C$. Again using Lemma 9.1, we may replace conditions (11.4), (11.5) by

$$\sum_{n_0 < \bar{n}} \|(A_{n_0}(\omega, 0) - E)^{-1}\|_{\mathrm{HS}}^2 \geq C^{2n} \tag{12.1}$$

and

$$\|M_n(\omega, \theta, E)\|_{\mathrm{HS}}^{2J} \leq e^{-2n^{1-\sigma}J} \prod_1^J \|M_n(\omega, j\omega, E)\|_{\mathrm{HS}}^2 \tag{12.2}$$

which are polynomial inequalities in $\cos\omega_\alpha, \sin\omega_\alpha, \cos\theta_\alpha, \sin\theta_\alpha (\alpha = 1, 2)$ and $E$ of degree at most $n^C$. Restricting $\omega$ to $[0, 1]^2$, our aim is to get instead polynomials

$$P_1(\omega_1, \omega_2, E) \geq 0, \tag{12.3}$$

$$P_2(\omega_1, \omega_2, \cos\theta_1, \sin\theta_1, \cos\theta_2, \sin\theta_2, E) \geq 0. \tag{12.4}$$

This may be achieved in a straightforward way, if we go back to (11.4), (11.5) and truncate the $e^{ik\omega_\alpha} (\alpha = 1, 2)$ power series expansions at degree $\sim \bar{n}$. The polynomials in (12.3), (12.4) resulting from inequalities (12.1), (12.2) are still of degree at most $n^C$ for some constant $C$. In the case of a real analytic $v$, one proceeds again by truncation of the Fourier expansion of $v$ to get a trigonometric polynomial.

We denote by

$$\mathcal{A} \subset [0,1]^2 \times \mathbb{T}^2 \times [E_1, E_2] \tag{12.5}$$

the set defined by (12.3), (12.4) (restricting $E$ to the range $[E_1, E_2]$); Lemma 11.1 asserts thus that

$$\text{mes}\,[P_{\omega,\theta}(\mathcal{A}) \cap (\Omega \times \mathbb{T}^2)] < e^{-n^\sigma} \tag{12.6}$$

where

$$\Omega = \{\omega \in [0, 2\pi]^2 \big|\ \|k\omega\| > c|k|^{-A} \ \text{ for } \ 0 \neq |k| < n^C\}. \tag{12.7}$$

## 13. Frequency estimates

Fix $N$ such that

$$\log n \ll \log N; \qquad \log\log N \ll \log n \tag{13.1}$$

We estimate

$$\text{mes}\,\{\omega \in \Omega |\, (\omega, \ell\omega) \in \mathrm{P}_{\omega,\theta}(\mathcal{A}) \text{ for some } \ell \sim \mathrm{N}\}. \tag{13.2}$$

Thus we consider again

$$\sum_{\ell\sim N} \int_\Omega \chi_{P_{\omega,\theta}(\mathcal{A})}(\omega_1, \omega_2, \ell\omega_1, \ell\omega_2) d\omega_1 d\omega_2 \tag{13.3}$$

$$\leq \frac{1}{N^2} \sum_{\substack{\ell\sim N \\ |m_1|,|m_2|\leq N}} \int \chi_{P_{\omega,\theta}(\mathcal{A})\cap(\Omega\times\mathbb{T}^2)} \left(\frac{\theta_1 + m_1}{\ell}, \frac{\theta_2 + m_2}{\ell}, \theta_1, \theta_2\right) d\theta. \tag{13.4}$$

By (12.6), we may restrict the $\theta$-integration to $\theta$'s for which

$$\text{mes}_\omega(P_{\omega,\theta}(\mathcal{A}) \cap \Omega) < e^{-\frac{1}{2}n^\sigma}. \tag{13.5}$$

Fix $\theta \in \mathbb{T}^2$ satisfying (13.5) and estimate

$$\frac{1}{N^2} \sum_{\substack{\ell\sim N \\ |m_1|,|m_2|\leq N}} \chi_{P_\omega(\mathcal{A}_\theta)\cap\Omega'} \left(\frac{\theta_1 + m_1}{\ell}, \frac{\theta_2 + m_2}{\ell}\right) \tag{13.6}$$

where

$$\Omega' = \{\omega \in [0,1]^2 |\ \|k\omega\| > 2c|k|^{-A} \text{ for } \ 0 \neq |k| < n^C\}. \tag{13.7}$$

Thus if $\omega' \in \Omega'$, $|\omega - \omega'| < e^{-\frac{1}{10}n^\sigma}$, then $\omega \in \Omega$. Hence (13.5) implies that if $\omega \in P_\omega(\mathcal{A}_\theta) \cap \Omega'$, then

$$\text{dist}\,(\omega, \partial P_\omega(\mathcal{A}_\theta)) < e^{-\frac{1}{10}n^\sigma}. \tag{13.8}$$

For fixed $\theta$, we specify the boundary

$$\partial P_\omega(\mathcal{A}_\theta). \tag{13.9}$$

From (12.3), (12.4), the set $P_\omega(\mathcal{A}_\theta)$ is the $\omega = (\omega_1, \omega_2)$-projection of a set

$$[P_1(\omega_1, \omega_2, E) \geq 0] \cap [P_3(\omega_1, \omega_2, E) \geq 0] \tag{13.10}$$

where $P_1, P_3 \in \mathbb{R}[\omega_1, \omega_2, E]$ are polynomials of degree $< n^C$, for some constant $C$ (we will again use the letter $C$ in the sequel for various constants).

Factor

$$P_1 = \prod_\alpha p_\alpha^{\pi_\alpha} \text{ and } P_3 = \prod_\beta p_\beta^{\pi'_\beta} \tag{13.11}$$

in irreducible components $p_\alpha \in \mathbb{R}[\omega_1, \omega_2, E]$.

Lemma.

$$\partial P_\omega(\mathcal{A}_\theta) \subset \bigcup_{\alpha \neq \beta} [\mathcal{R}_E(p_\alpha, p_\beta) = 0] \cup \bigcup_\alpha [\mathcal{R}_E(p_\alpha, \partial_E p_\alpha) = 0] \tag{13.12}$$

*where* $\mathcal{R}_E(p_\alpha, p_\beta) = \mathcal{R}_E(p_\alpha, p_\beta)(\omega_1, \omega_2)$ *denotes the resultant of* $p_\alpha, p_\beta \in \mathcal{R}[\omega_1, \omega_2][E]$.

*Proof of the lemma.* Let $\omega = (\omega_1, \omega_2) \in \partial P_\omega(\mathcal{A}_\theta)$ and $(\omega, E) \in \mathcal{A}_\theta$. Clearly $p_\alpha(\omega, E) = 0$ for some $\alpha$. If $p_\beta(\omega, E) = 0$ for some $\beta \neq \alpha$, then $p_\alpha(\omega, \cdot)$ and $p_\beta(\omega, \cdot)$ have a common root and $\mathcal{R}_E(p_\alpha, p_\beta) = 0$. Assume now $p_\beta(\omega, E) \neq 0$ for all $\beta \neq \alpha$. Thus for $\beta \neq \alpha$, $p_\beta(\omega, E)$ and $p_\beta(\omega', E')$ will have the same sign, if $(\omega', E')$ is close enough to $(\omega, E)$. Since $\mathcal{R}_E(p_\alpha, \partial_E p_\alpha) \neq 0$, $\partial_E p_\alpha(\omega, E) \neq 0$. One may then find $E', E''$ near $E$ such that $p_\alpha(\omega, E') > 0$, $p_\alpha(\omega, E'') < 0$. Hence, also $p_\alpha(\omega', E') > 0$, $p_\alpha(\omega', E'') < 0$ for $\omega'$ near $\omega$ and $p_\alpha(\omega', E_{\omega'}) = 0$ for some $E_{\omega'} \in [E', E'']$. For $\beta \neq \alpha$, we have that sign $p_\beta(\omega', E_{\omega'}) =$ sign $p_\beta(\omega, E)$.

Consider the polynomial $P_1$. If $p_\alpha$ is not a factor of $P_1$, then $P_1(\omega, E) > 0$ and $P_1(\omega', E_{\omega'}) > 0$. Otherwise, $P_1(\omega', E_{\omega'}) = 0$. In both cases $P_1(\omega', E_{\omega'}) \geq 0$. The same holds for $P_3$. It follows that $(\omega', E_{\omega'}) \in \mathcal{A}_\theta$; thus $\omega' \in P_\omega(\mathcal{A}_\theta)$ for all $\omega'$ near $\omega$. Therefore $\omega \notin \partial P_\omega(\mathcal{A}_\theta)$. This proves the lemma.

From the irreducibility assumption

$$\mathcal{R}_E(p_\alpha, p_\beta)(\omega_1, \omega_2) \not\equiv 0 \text{ for } \alpha \neq \beta \tag{13.13}$$

and

$$\mathcal{R}_E(p_\alpha, \partial_E p_\alpha)(\omega_1, \omega_2) \not\equiv 0. \tag{13.14}$$

Thus $\partial P_\omega(\mathcal{A}_\theta)$ is contained in the union of at most $n^C$ (irreducible) algebraic curves

$$\Gamma = [\omega | P(\omega) = 0] \tag{13.15}$$

where $P(\omega) \in \mathbb{R}[\omega_1, \omega_2]$ is an irreducible polynomial of degree $< n^C$.

From (13.8), a bound on (13.6) will thus result from the estimations

$$\#\{\ell \sim N, |m_1|,\ m_2| \leq N \Big| \frac{\theta+m}{\ell} \in P_\omega(\mathcal{A}_\theta) \cap \Omega' \text{ and } \operatorname{dist}\Big(\frac{\theta+m}{\ell}, \Gamma\Big) < e^{-\frac{1}{10}n^\sigma}\} \tag{13.16}$$

with $\Gamma$ of the form (13.15); the bound on (13.16) needs then to be multiplied by $n^C.N^{-2}$.

Fix $\delta > 0$ such that

$$\frac{1}{\delta} < N, \ \log\frac{1}{\delta} \gg \log n \tag{13.17}$$

and subdivide $\Omega' \subset [0,1]^2$ in squares $I$ of size $\delta$. Since $\Gamma$ is a union of at most $n^C$ connected components and, by the integral formula, $\ell(\Gamma) < n^C$, the number of $I$-squares intersecting $\Gamma$ is at most $\delta^{-1}n^C$.

Assume

$$(13.16) > \kappa N^2. \tag{13.18}$$

Fixing $\ell \sim N$, assume that

$$\#\Big\{|m_1|, |m_2| \leq N \Big| \frac{\theta+m}{\ell} \in P_\omega(\mathcal{A}_\theta) \cap \Omega',\ \operatorname{dist}\Big(\frac{\theta+m}{\ell}, \Gamma\Big) < e^{-\frac{1}{10}n^\sigma}\Big\} > \kappa N. \tag{13.19}$$

Using the previous covering of $\Gamma$ by $\delta$-squares $I$, assume

$$\#\Big\{|m_1|, |m_2| \leq N \Big| \frac{\theta+m}{\ell} \in P_\omega(\mathcal{A}_\theta) \cap I,\ \operatorname{dist}\Big(\frac{\theta+m}{\ell}, \Gamma\Big) < e^{-\frac{1}{10}n^\sigma}\Big\} > n^{-C}\kappa\delta N. \tag{13.20}$$

Observe that in (13.20) $m_1, m_2$ vary in intervals of size $\delta N$. Hence, trivially

$$(13.20) < (\delta N)^2. \tag{13.21}$$

Assume also

$$\nabla P \neq 0 \ \text{ on } \ I \cap \Gamma. \tag{13.22}$$

Since the number of singular points of $\Gamma$ is $< n^C$, this excludes at most $n^C$ squares, with contribution

$$< n^C(\delta N)^2 \tag{13.23}$$

in (13.19).

Since the number of components of $\Gamma \cap I$ is bounded by $n^C$, we may in (13.20) replace $\Gamma$ by $\Gamma_0$ satisfying

$$\Gamma_0 \subset \Gamma \cap I,\ \Gamma_0 \text{ is connected.} \tag{13.24}$$

Assume

$$\omega^i = \frac{\theta + m^i}{\ell}\ (i = 1, 2, 3) \tag{13.25}$$

three noncolinear points in the set ($\ell$ fixed)

$$\left\{ \omega = \frac{\theta + m}{\ell} \in P_\omega(\mathcal{A}_\theta) \cap I \Big| \operatorname{dist}(\omega, \Gamma_0) < e^{-\frac{1}{10}n^\sigma} \right\}. \tag{13.26}$$

Let then

$$\bar{\omega}^i \in \Gamma_0,\ |\omega^i - \bar{\omega}^i| < e^{-\frac{1}{10}n^\sigma}. \tag{13.27}$$

Since by assumption

$$|\det(\omega^1 - \omega^0, \omega^2 - \omega^0)| = \left| \det\left( \frac{m^1 - m^0}{\ell}, \frac{m^2 - m^0}{\ell} \right) \right| \geq \frac{1}{\ell^2} \tag{13.28}$$

and $\log\log N \ll \log n$, (13.27), it follows that

$$\det(\bar{\omega}^1 - \bar{\omega}^0, \bar{\omega}^2 - \bar{\omega}^0) > \frac{1}{2\ell^2}. \tag{13.29}$$

Denote by $\gamma_1 = \gamma_1(s), s \in [0, 1]$ (resp. $\gamma_2$), continuous curves contained in $\Gamma_0$ such that

$$\begin{aligned} &\gamma_1(0) = \bar{\omega}^0,\ \gamma_1(1) = \bar{\omega}^1, \\ &\gamma_2(0) = \bar{\omega}^0,\ \gamma_2(1) = \bar{\omega}^2 \end{aligned} \tag{13.30}$$

and

$$\operatorname{Var}\gamma_\alpha \leq \ell(\Gamma_0) < n^C\delta \qquad (\alpha = 1, 2). \tag{13.31}$$

Let $\varepsilon \overset{>}{\to} 0$ and write from (13.29)

$$\begin{aligned} \frac{1}{2\ell^2} &< \det(\gamma_1(1) - \gamma_1(0),\ \gamma_2(1) - \gamma_2(0)) \\ &= \sum_{k_1, k_2 = 0}^{1/\varepsilon} \det[\gamma_1((k_1 + 1)\varepsilon) - \gamma_1(k_1\varepsilon), \gamma_2((k_2 + 1)\varepsilon) - \gamma_2(k_2\varepsilon)] \end{aligned} \tag{13.32}$$

while, by (13.31)

$$\sum_k |\gamma_\alpha((k + 1)\varepsilon) - \gamma_\alpha(k\varepsilon)| < n^C\delta. \tag{13.33}$$

Hence, from (13.32), (13.33), there are $1 \leq k_1, k_2 \leq \frac{1}{\varepsilon}$ satisfying

(13.34)
$$|\det[\gamma_1\big((k_1+1)\varepsilon\big) - \gamma_1(k_1\varepsilon), \gamma_2\big((k_2+1)\varepsilon\big) - \gamma_2(k_2\varepsilon)]| > n^{-C}\delta^{-2}\ell^{-2}|\gamma_1\big((k_1+1)\varepsilon\big) - \gamma_1(k_1\varepsilon)| \cdot |\gamma_2\big((k_2+1)\varepsilon\big) - \gamma_2(k_2\varepsilon)|.$$

Also, for $\gamma = \gamma_1$, or $\gamma_2$

$$P\big(\gamma(k\varepsilon)\big) = 0; \tag{13.35}$$

hence

$$\begin{aligned} 0 &= P\big(\gamma\big((k+1)\varepsilon\big)\big) - P\big(\gamma(k\varepsilon)\big) \\ &= (\nabla P)\big(\gamma(k\varepsilon)\big) \cdot [\gamma\big((k+1)\varepsilon\big) - \gamma(k\varepsilon)] + O\big(|\gamma\big((k+1)\varepsilon\big) - \gamma(k\varepsilon)|^2\big) \end{aligned} \tag{13.36}$$

and, from the critical point assumption,

$$\frac{\nabla P}{|\nabla P|}\big(\gamma(k\varepsilon)\big) \cdot \frac{\gamma\big((k+1)\varepsilon\big) - \gamma(k\varepsilon)}{|\gamma\big((k+1)\varepsilon\big) - \gamma(k\varepsilon)|} = \varepsilon'(\varepsilon) \overset{\varepsilon \to 0}{\to} 0. \tag{13.37}$$

Properties (13.34), (13.37) imply that

$$\begin{aligned} &\text{angle } \left(\frac{\gamma_1\big((k_1+1)\varepsilon\big) - \gamma_1(k_1\varepsilon)}{|\gamma_1\big((k_1+1)\varepsilon\big) - \gamma_1(k_1\varepsilon)|}, \frac{\gamma_2\big((k_2+1)\varepsilon\big) - \gamma_2(k_2\varepsilon)}{|\gamma_2\big((k_2+1)\big) - \gamma_2(k_2\varepsilon)|}\right) \\ &\approx \text{angle}\left(\frac{\nabla P}{|\nabla P|}\big(\gamma_1(k_1\varepsilon)\big), \frac{\nabla P}{|\nabla P|}\big(\gamma_2(k_2\varepsilon)\big)\right) > n^{-C}\delta^{-2}\ell^{-2}. \end{aligned} \tag{13.38}$$

Consequently the map

$$\frac{\nabla P}{|\nabla P|} : \Gamma_0 \to S^1 \tag{13.39}$$

covers an arc of size $n^{-C}\delta^{-2}\ell^{-2}$.

Suppose we get $M$ squares $I$ such that (13.26) contains three noncolinear points. Thus for each of these $I$'s, $\frac{\nabla P}{|\nabla P|}(\Gamma \cap I)$ covers an arc in $S^1$ of size $n^{-C}\delta^{-2}\ell^{-2}$. Hence, if

$$J = Mn^{-C}\delta^{-2}\ell^{-2} \gg 1 \tag{13.40}$$

we may find distinct points $(\omega^j)_{1\leq j \leq J}$ on $\Gamma$ such that

$$\nabla P(\omega^j) \neq 0, \ \frac{\nabla P}{|\nabla P|}(\omega^j) = \zeta \qquad (1 \leq j \leq J) \tag{13.41}$$

for some $\zeta \in S^1$. We may moreover assume

$$\langle \zeta^\perp, \nabla P \rangle \not\equiv 0. \tag{13.42}$$

Since, by (13.15), (13.41),

$$\#([P=0]\cap[\langle\zeta^{\perp},\nabla P\rangle=0])\geq J \tag{13.43}$$

and $P$ is irreducible of degree $< n^C$, it follows from Bezout's theorem that $J < n^C$. Hence

$$M < n^C\delta^2\ell^2. \tag{13.44}$$

By (13.21), the contribution of those $I$'s for which (13.26) contains at least three noncolinear points is thus at most

$$n^C(\delta N)^4. \tag{13.45}$$

Assume next that the points in (13.26) are colinear. Thus there is a line $L$ such that ($\ell$ fixed)

$$\#\{\omega=\frac{\theta+m}{\ell}\in P_\omega(\mathcal{A}_\theta)\cap I\cap L\} > n^{-C}\kappa\delta N. \tag{13.46}$$

Observe that the translated line $0\in L_0//L$ necessarily contains an element $m=m_I\in\mathbb{Z}^2\backslash\{0\}$ satisfying

$$|m_I| < n^C\kappa^{-1}. \tag{13.47}$$

Since $P_\omega(\mathcal{A}_\theta)\cap I\cap L$ has at most $n^C$ components, it follows that $P_\omega(\mathcal{A}_\theta)\cap I\cap L$ contains an interval of size $> n^{-C}\kappa\delta$, provided we assume

$$\kappa\delta N > n^C. \tag{13.48}$$

Assume (cf. (13.45))

$$n^C(\delta N)^4\ll\kappa N;$$

i.e.,

$$\delta^4N^3\ll n^{-C}\kappa. \tag{13.49}$$

By (13.19) and since the left side of (13.46) is at most $\delta N$, we obtain at least $\kappa\delta^{-1}$ squares $I$, from the $n^C\delta^{-1}$ intersecting $\Gamma$, that have the previous line property. Next, we let $\ell$ vary. Since for fixed $\ell$, (13.19) is at most $n^C\delta^{-1}(\delta N)^2=n^C\delta N^2$, (13.18) implies that (13.19) needs to hold for at least $\kappa n^{-C}\delta^{-1}$ values of $\ell$. For each of these, we get $\kappa\delta^{-1}$ squares $I$ with the line property obtained above. Since the total number of $\delta$-squares intersecting $\Gamma$ is at most $n^C\delta^{-1}$, there is some $\delta$-square $I$ such that (13.46) holds for at least $n^{-C}\kappa^2\delta^{-1}$ values of $\ell$. Moreover, by (13.47), we may assume those lines parallel for at least $n^{-C}\kappa^4\delta^{-1}$ values of $\ell$. Thus for each of these $\ell$, there is a line $L_\ell$ such that

$$L_\ell//\bar{m}\in\mathbb{Z}^2\backslash\{0\}, \tag{13.50}$$

$$L_\ell\cap I\cap P_\omega(\mathcal{A}_\theta)\text{ contains a segment of size } > n^{-C}\kappa\delta. \tag{13.51}$$

Our purpose is to use now the fact that

$$\text{mes}\,(P_\omega(\mathcal{A}_\theta)\cap I) \leq \text{mes}\,(P_\omega(\mathcal{A}_\theta)\cap \Omega) < e^{-\frac{1}{2}n^\sigma} \tag{13.52}$$

according to (13.5).

First, estimate from below $\text{dist}\,(L_{\ell_1}, L_{\ell_2})$ for $\ell_1 \neq \ell_2$. The line $L_\ell$ is parametrized as

$$\frac{\theta + m^{(\ell)} + t\bar{m}}{\ell} \ (t\in\mathbb{R}) \text{ for some } m^{(\ell)} \in \mathbb{Z}^2. \tag{13.53}$$

Hence, since $|\bar{m}| < N$

$$\text{dist}\,(L_{\ell_1}, L_{\ell_2}) > \frac{1}{N^3}\|(\ell_1-\ell_2)\bar{m}_2\theta_1 - (\ell_1-\ell_2)\bar{m}_1\theta_2\| > N^{-3}\gamma_\theta \tag{13.54}$$

where $\gamma_\theta$ is defined by

$$\|k.\theta\| > \gamma_\theta \ \text{ for } \ k\in\mathbb{Z}^2\backslash\{0\}, |k| < N^2. \tag{13.55}$$

Assume $\theta$ such that

$$\gamma_\theta > e^{-\frac{1}{100}n^\sigma}. \tag{13.56}$$

The contribution in (13.4) of those $\theta$'s for which $\gamma_\theta \leq e^{-\frac{1}{100}n^\sigma}$ is by (13.55) clearly bounded by $N^4 e^{-\frac{1}{100}n^\sigma} < e^{-\frac{1}{200}n^\sigma}$.

Denote

$$e_1 = \frac{\bar{m}}{|\bar{m}|} \text{ and } e_2 = e_1^\perp = \frac{(\bar{m}_2, -\bar{m}_1)}{|\bar{m}|}. \tag{13.57}$$

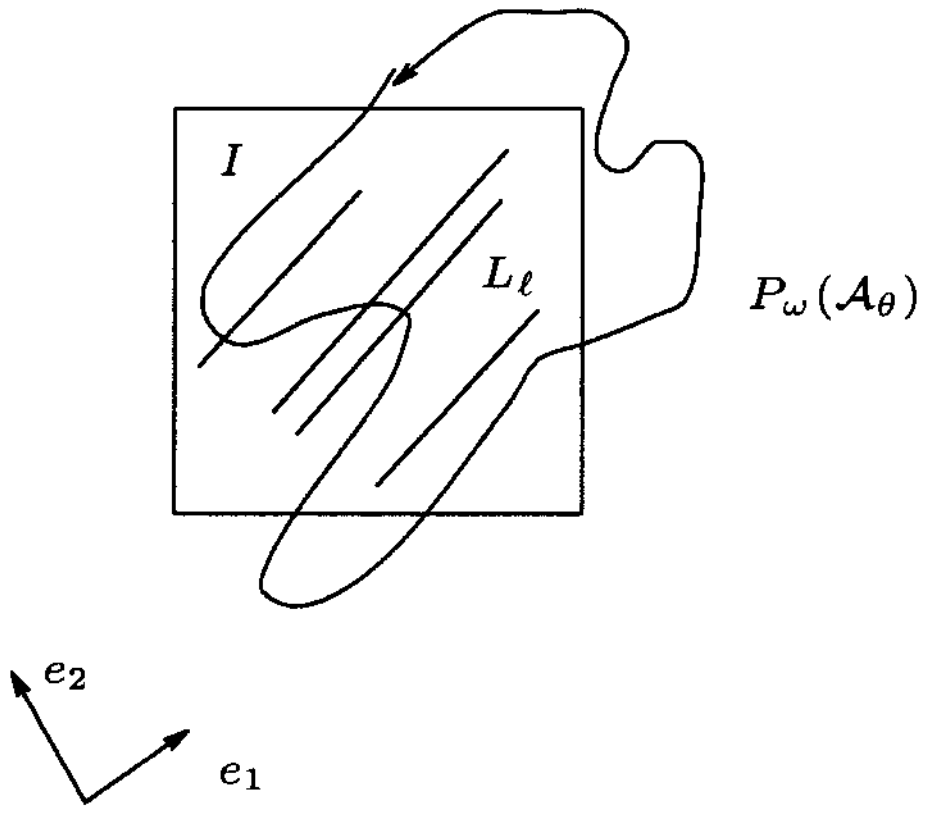

From the preceding, we have

$$\begin{aligned} n^{-C}\kappa^5 &< \sum_{\ell\sim N} P_{e_1}(P_\omega(\mathcal{A}_\theta)\cap I\cap L_\ell) \\ &= \int_{P_{e_1}(I)} [\#\{\ell\sim N|\ (te_1+\mathbb{R}e_2)\cap P_\omega(\mathcal{A}_\theta)\cap I\cap L_\ell \neq \emptyset\}]dt. \end{aligned} \tag{13.58}$$

Since $(te_1 + \mathbb{R}e_2) \cap P_\omega(\mathcal{A}_\theta) \cap I$ is a union of at most $n^C$ segments, it follows from (13.54) that

$$\#\{\ell \sim N | (te_1 + \mathbb{R}e_2) \cap P_\omega(\mathcal{A}_\theta) \cap I \cap L_\ell \neq \emptyset\} \leq n^C + N^3 \gamma_\theta^{-1} |(te_1 + \mathbb{R}e_2) \cap P_\omega(\mathcal{A}_\theta) \cap I|. \tag{13.59}$$

Therefore (13.58), (13.59), (13.52), (13.56) imply

$$\begin{aligned} n^{-C}\kappa^5 &< n^C\delta + N^3\gamma_\theta^{-1}|P_\omega(\mathcal{A}_\theta) \cap I| \\ &< n^C\delta + N^3 e^{\frac{1}{100}n^\sigma} e^{-\frac{1}{2}n^\sigma} \\ &< n^C\delta + e^{-\frac{1}{3}n^\sigma}. \end{aligned} \tag{13.60}$$

Hence

$$\kappa < n^C\delta^{1/5}. \tag{13.61}$$

Recalling conditions (13.17), (13.48), (13.49) we see that

$$\begin{cases} \log\log N \ll \log n \ll \log N \\ \frac{1}{\delta} < N, \log\frac{1}{\delta} \gg \log n \\ \kappa\delta N > n^C \\ \delta^4 N^3 \ll n^{-C}\kappa. \end{cases} \tag{13.62}$$

When

$$\delta = N^{-7/8}, \tag{13.63}$$

(13.61) is not compatible with $\kappa\delta N > n^C$ in (13.62). Therefore

$$\kappa < n^C N^{-\frac{1}{8}} < N^{-\frac{1}{9}}. \tag{13.64}$$

Thus, by (13.18),

$$(13.16) < N^{2-\frac{1}{9}} \tag{13.65}$$

except for a set of $\theta$'s contributing in (13.4) at most $e^{-\frac{1}{200}n^\sigma}$. Consequently

$$(13.4) < n^C N^{-\frac{1}{9}} \tag{13.66}$$

and

$$(13.2) \leq \sum_{\ell \sim N} \int_\Omega \chi_{P_{\omega,\theta}(\mathcal{A})}(\omega, \ell\omega) d\omega < N^{-\frac{1}{10}}. \tag{13.67}$$

The conclusion is the following:

Lemma 13.68. *Choose $\delta > 0$ and $n$ a sufficiently large integer. Denote by $\Omega \subset \mathbb{T}^2$ the set of frequencies $\omega$ such that*

$$\omega \in DC_{10,c}, \tag{13.69}$$

(13.70) *There is* $n_0 < n^C$, $\quad 2^{(\log n)^2} \leq \ell \leq 2^{(\log n)^3}$ *and* $E$ *such that*

$$\|[A_{n_0}(\omega, 0) - E]^{-1}\| > C^n, \tag{13.71}$$

$$\frac{1}{n}\log\|M_n(\omega, \ell\omega, E)\| < L_n(\omega, E) - \delta. \tag{13.72}$$

*Then*

$$\operatorname{mes}\Omega < e^{-\frac{1}{20}(\log n)^2}.$$

The proof of Theorem 1 for $d = 2$ may then be completed as in the 1-frequency case.

## VI. Appendix: Lyapounov exponents

### 14. The one-variable case

In this section, we give an alternate proof of the Sorets-Spencer result [S-S] (Prop. 14.8 below). Let $v$ be a 1-periodic real analytic function on $\mathbb{R}$:

$$v(\theta) = \sum_{k\in\mathbb{Z}} \widehat{v}(k) e^{2\pi i k\theta}, \tag{14.1}$$

$$|\widehat{v}(k)| \lesssim e^{-\rho|k|} \tag{14.2}$$

for some $\rho > 0$. Thus there is a holomorphic extension

$$v(z) = \sum_{k\in\mathbb{Z}} \widehat{v}(k) e^{2\pi i k z} \tag{14.3}$$

to the strip $|\operatorname{Im} z| < \frac{\rho}{10}$, satisfying

$$|v(z)| \leq \sum_{k\in\mathbb{Z}} |\widehat{v}(k)| e^{2\pi|k|\ |\operatorname{Im} z|} < \sum_{k\in\mathbb{Z}} e^{-\rho|k|} e^{\rho|k|\frac{\pi}{5}} < C. \tag{14.4}$$

LEMMA 14.5. *Assume* $v$ *is not constant. For all* $0 < \delta < \rho$, *there is* $\varepsilon$ *such that*

$$\inf_{E_1} \sup_{\frac{\delta}{2}<y<\delta} \inf_{x\in[0,1]} |v(x + iy) - E_1| > \varepsilon. \tag{14.6}$$

*Proof.* Assume (14.6) fails. From a compactness argument, it follows that there is $E_1$ and for all $\frac{\delta}{2} < y < \delta$ some $0 \leq x(y) \leq 1$ such that

$$v\big(x(y) + iy\big) - E_1 = 0. \tag{14.7}$$

Thus $v(z) - E_1$ would have infinitely many zeros on $[0, 1] \times [0, \delta]$, implying $v \equiv E_1$ (contradiction).

Next, we give a proof of the Sorets-Spencer result:

PROPOSITION 14.8 ([S-S]). *If $v$ is as in Lemma 14.5, there is $\lambda_0 > 0$ such that for all $E$ and $\lambda > \lambda_0$, the Lyapounov exponent of $\lambda v$,*

$$L(\omega, E) = \inf_n \frac{1}{n} \int_0^1 \log \|M_n(\omega, \theta, E)\| d\theta > c \log \lambda \tag{14.9}$$

*where* $M_n(\omega, \theta, E) = \prod_{j=n}^{1} \begin{pmatrix} \lambda v(\theta + j\omega) - E & -1 \\ 1 & 0 \end{pmatrix}$ *and* $c = c_v$ *a constant.*

*Proof.* By (14.4), the function

$$M_n(\omega, z, E) = \prod_{j=n}^{1} \begin{pmatrix} \lambda v(z + j\omega) - E & -1 \\ 1 & 0 \end{pmatrix} \tag{14.10}$$

is analytic on $|\mathrm{Im}\, z| < \frac{\rho}{10}$, bounded by $(C\lambda + E + 1)^n < (C\lambda)^n$ (we assume $|E| < C\lambda$, which is clearly no restriction).

Hence

$$\varphi(z) = \frac{1}{n} \log \|M_n(\omega, z, E)\| \tag{14.11}$$

is a subharmonic function on $|\mathrm{Im}\, z| < \frac{\rho}{10}$, bounded by

$$|\varphi(z)| < \log C\lambda \quad \text{for } |\mathrm{Im}\, z| < \frac{\rho}{10}. \tag{14.12}$$

Fix $0 < \delta \ll \rho$ and $\varepsilon$ satisfying Lemma 14.5. Define

$$\lambda_0 = 100\, \varepsilon^{-100} \tag{14.13}$$

and let $\lambda > \lambda_0$. With $E$ fixed, there is thus $\frac{\delta}{2} < y_0 < \delta$ such that

$$\inf_{x \in [0,1]} \left| v(x + iy_0) - \frac{E}{\lambda} \right| > \varepsilon; \tag{14.14}$$

hence, since $v$ is 1-periodic $v(z) = v(z+1)$

$$\inf_{x \in \mathbb{R}} |\lambda v(x + iy_0) - E| > \lambda\varepsilon > 100. \tag{14.15}$$

Returning to (14.10), we claim that

$$\|M_n(\omega, iy_0, E)\| > (\lambda\varepsilon - 1)^n. \tag{14.16}$$

To see (14.16), write

$$M_{n-1}(\omega, iy_0, E) \begin{pmatrix} 1 \\ 0 \end{pmatrix} = \begin{pmatrix} u_{n-1} \\ v_{n-1} \end{pmatrix}. \tag{14.17}$$

Thus,

$$\binom{u_n}{v_n} = \begin{pmatrix} \lambda v(iy_0 + n\omega) - E & -1 \\ 1 & 0 \end{pmatrix} \binom{u_{n-1}}{v_{n-1}} = \binom{[\lambda v(iy_0 + n\omega) - E]u_{n-1-v_{n-1}}}{u_{n-1}}. \tag{14.18}$$

Hence, by (14.15),

$$|u_n| > \lambda\varepsilon |u_{n-1}| - |v_{n-1}| > 100|u_{n-1}| - |v_{n-1}|; |v_n| = |u_{n-1}| \tag{14.19}$$

from which we deduce that

$$|u_n| \geq |v_n| \tag{14.20}$$

and

$$|u_n| > (\lambda\varepsilon - 1)|u_{n-1}| > (\lambda\varepsilon - 1)^n, \tag{14.21}$$

implying (14.16).

Consequently

$$\varphi(iy_0) > \log(\lambda\varepsilon - 1). \tag{14.22}$$

Denote $\mu \in M([\operatorname{Im} z = 0] \cup [\operatorname{Im} z = \frac{\rho}{10}])$ the harmonic measure of $y_0$ in the strip

$$0 \leq \operatorname{Im} z \leq \frac{\rho}{10}. \tag{14.23}$$

Clearly

$$\mu\Big(\Big[\operatorname{Im} z = \frac{\rho}{10}\Big]\Big) < \frac{10\delta}{\rho} \text{ and } \frac{d\mu}{dx}\Big|_{\operatorname{Im} z = 0} \leq \frac{y_0}{y_0^2 + x^2}. \tag{14.24}$$

It follows from (14.12), (14.22), (14.24) that

(14.25)

$$\begin{aligned} \log(\lambda\varepsilon - 1) < \varphi(iy_0) &< \int_{[y=0]\cup[y=\frac{\rho}{10}]} \varphi(z)\mu(dz) \qquad (z = x + iy) \\ &< \int_{-\infty}^{\infty} \varphi(x) \frac{y_0}{y_0^2 + x^2} dx + \frac{C\delta}{\rho} \log C\lambda \\ &\leq \int_0^1 \varphi(\theta) \Big( \sum_{k\in\mathbb{Z}} \frac{y_0}{y_0^2 + (\theta + k)^2} \Big) d\theta + \frac{C\delta}{\rho} \log \lambda. \end{aligned}$$

Thus

(14.26)

$$\frac{1}{n}\int_0^1 \log \|M_n(\omega,\theta,E)\| d\theta = \int_0^1 \varphi(\theta)d\theta \geq \frac{y_0}{2}\int_0^1 \varphi(\theta)\Big(\sum_{k\in\mathbb{Z}} \frac{y_0}{y_0^2+(\theta+k)^2}\Big)d\theta$$

$$\overset{(14.25)}{\geq} \frac{y_0}{2}\Big(\log(\lambda\varepsilon-1) - \frac{C\delta}{\rho}\log\lambda\Big)$$

$$> \frac{\delta}{4}\Big(\Big(1-\frac{C\delta}{\rho}\Big)\log\lambda - 2\log\frac{1}{\varepsilon}\Big).$$

Since $\delta \ll \rho$, and by (14.13),

$$(14.26) > \frac{\delta}{4}\Big(\frac{1}{2}\log\lambda - \frac{1}{50}\log\lambda_0\Big) > \frac{\delta}{16}\log\lambda \tag{14.27}$$

proving Proposition 14.8.

## 15. The higher dimensional case

The previous argument based on complexification does not seem to apply directly in the case of real analytic $v$ on $\mathbb{T}^d, d > 1$ (except in special cases). We will develop here a different approach, mainly based on the use of the large deviation theorem (Lemma 8.1) and the resolvent identity. The purpose of what follows is to establish a recursive inequality relating the numbers $L_n = L_n(\omega, E)$ $\big($see (0.6)$\big)$ for different values of $n$. Let thus $v$ be a real analytic potential on $\mathbb{T}^d$ and let $\|v\|_\infty = \sup_{|\mathrm{Im}\, z_j|<\rho} |v(z_1,\dots,z_d)|$ where $v(z_1,\dots,z_d) = \sum_{k\in\mathbb{Z}^d} \hat{v}(k)e^{ik.z}$ is the holomorphic extension of $v$ to some strip $[|\mathrm{Im}\, z_j| < \rho; 1 \leq j \leq d]$ $(\rho > 0)$. Let $\omega \in \mathbb{T}^d$ be a fixed diophantine frequency vector. We will denote $M_n = M_n(\theta) = M_n(\omega,\theta,E)$ and $L_n = L_n(\omega,E)$ for simplicity. Thus

$$\frac{1}{n}\log\|M_n\| \leq \log(1+\|v\|_\infty), \tag{15.1}$$

$$L_n \leq \log(1+\|v\|_\infty). \tag{15.2}$$

*Step* 1.

LEMMA 15.3. *Fix a large integer $n$ and assume*

$$\rho = \frac{\log(1+\|v\|_\infty)}{L_n}(\log n)^{-1/2} < \frac{1}{2}. \tag{15.4}$$

*There is an integer $n_0$ such that*

$$n^{1/2} < n_0 \leq n \tag{15.5}$$

*and*

$$L_m < (1+\rho)L_{n_0} + \log(1+\|v\|_\infty)\cdot\frac{\rho n_0}{m} \tag{15.6}$$

*whenever*

$$m \le n_0. \tag{15.7}$$

*Proof.* By (0.18), it suffices to establish the inequality

$$L_{[\rho n_0]} < (1+\rho)L_{n_0} \tag{15.8}$$

for some $n^{1/2} < n_0 \le n$. If the property fails for $n$, consider $n_1 = [\rho n]$ and so on. This generates a sequence

$$n_{j+1} = \rho n_j + 0(1) \tag{15.9}$$

where

$$L_{n_{j+1}} \ge (1+\rho)L_{n_j}. \tag{15.10}$$

Hence, by (15.2), (15.10),

$$\log(1+\|v\|_\infty) \ge L_{n_j} \ge (1+\rho)^j L_n \tag{15.11}$$

implying that

$$\begin{aligned} n_j > \rho^j n &> \frac{n}{\exp[\frac{1}{\rho}(\log\frac{1}{\rho})\cdot\log\frac{\log(1+\|v\|_\infty)}{L_n}]} \\ &> n.\exp[-(\log n)^{1/2}\log\log n] > n^{1/2}. \end{aligned} \tag{15.12}$$

The proof of the lemma is now clear.

*Remark.* The function $v$ will in fact be $v = \lambda v_0, v_0$ given, and $\lambda$ a large factor. It is therefore important to keep track of the effect of $\lambda \sim \|v\|_\infty$ in the various inequalities.

*Step* 2. Apply the large deviation estimate (Lemma 8.1). As we observed earlier, the result remains valid in the real analytic case and for arbitrary dimension $d$. Taking the previous remark into account, the function $\varphi$ given by (0.4) needs to be normalized by $[\log(1+\|v\|_\infty)]^{-1}$. The conclusion is that

$$\mathrm{mes}\,[\theta\in\mathbb{T}^{\mathrm{d}}|\ |\frac{1}{\mathrm{m}}\log\|\mathrm{M_m}(\theta)\| - \mathrm{L_m}| > \mathrm{m}^{-\sigma}\log(1+\|\mathrm{v}\|_\infty)] \lesssim \mathrm{e}^{-\mathrm{m}^\sigma}. \tag{15.13}$$

The constant $\sigma > 0$ depends only on $\omega$.

Define the next scale

$$N = [e^{n^{\sigma/10}}]. \tag{15.14}$$

Taking $n_0^{1/2} < m \le n_0$ in (15.13) we get thus a set $S \subset \mathbb{T}^d$ satisfying

(15.15) If $\theta \notin S$ and $0 \leq j \leq 2N$, then for $n_0^{1/2} < m \leq n_0$,

$$\left|\frac{1}{m}\log\|M_m(\theta + j\omega)\| - L_m\right| < n_0^{-\sigma/2}\log(1 + \|v\|_\infty) \tag{15.16}$$

and

$$\mathrm{mes\, S} < \sum_{\mathrm{n}_0^{1/2}\leq \mathrm{m} \leq \mathrm{n}_0} \mathrm{e}^{\mathrm{n}^{\sigma/10}}\, \mathrm{e}^{-\mathrm{m}^{\sigma}} \overset{(15.5)}{<} \mathrm{e}^{-\mathrm{n}^{\sigma/5}}. \tag{15.17}$$

Let $\theta \notin S$, $0 \leq j \leq N$ and denote $\theta' = \theta + j\omega$. By (0.3), there is

$$A' \in \{A_{n_0}(\theta'),\, A_{n_0-1}(\theta'),\, A_{n_0-1}(\theta' + \omega), A_{n_0-2}(\theta' + \omega)\} \tag{15.18}$$

such that

$$\log|\det(A' - E)| = \log\|M_{n_0}(\theta')\| + 0(1). \tag{15.19}$$

Hence, by (15.15),

$$\log|\det(A' - E)| \geq n_o L_{n_0} - n_0^{1-\frac{\sigma}{2}}\log(1 + \|v\|_\infty). \tag{15.20}$$

Recall also (0.22), for $1 \leq k_1 \leq k_2 \leq n_0$,

$$|(A' - E)^{-1}(k_1, k_2)| \leq \frac{\|M_{k_1-1}\big(\theta'(+\omega)\big)\|\ \|M_{n_0-k_2(-1,2)}\big(\theta' + k_2\omega(+\omega)\big)\|}{|\det(A' - E)|}. \tag{15.21}$$

Consider the numerator factors in (15.21) with argument

$$\theta'' \in \{\theta', \theta' + \omega, \theta' + k_2\omega, \theta' + (k_2 + 1)\omega\}.$$

For $m \leq n_0^{1/2}$, use the trivial bound

$$\|M_m(\theta'')\| < e^{n_0^{1/2}\log(1+\|v\|_\infty)}. \tag{15.22}$$

For $n_0^{1/2} < m \leq n_0$, (15.16) gives

$$\|M_m(\theta'')\| < e^{mL_m + n_0^{1-\frac{\sigma}{2}}\log(1+\|v\|_\infty)} \tag{15.23}$$

and (15.23) is thus valid in either case.

From (15.6), we deduce further that

$$\begin{aligned}\|M_m(\theta'')\| &< e^{(1+\rho)mL_{n_0} + 2\rho n_0\log(1+\|v\|_\infty)} \\ &< e^{mL_{n_0} + 3\rho n_0\log(1+\|v\|_\infty)}\end{aligned} \tag{15.24}$$

where $\rho$ is given by (15.4).

Substitution of (15.20), (15.24) in (15.21) implies thus that

$$\begin{aligned}|(A' - E)^{-1}(k_1, k_2)| &\leq e^{(n_0 - |k_1-k_2|)L_{n_0} + 6\rho n_0\log(1+\|v\|_\infty) - n_0 L_{n_0} + n_0^{1-\frac{\sigma}{2}}\log\|v\|_\infty} \\ &< e^{-|k_1-k_2|L_{n_0} + 7\rho n_0\log(1+\|v\|_\infty)}.\end{aligned} \tag{15.25}$$

It follows, in particular, from (15.25) that

$$|(A'-E)^{-1}(k_1,k_2)| < e^{-\left(L_{n_0}-70\rho\log(1+\|v\|_\infty)\right)|k_1-k_2|} \text{ if } |k_1-k_2| > \frac{n_0}{10}. \tag{15.26}$$

*Step* 3 (use of the resolvent identity). Fix $\theta \notin S$ and consider the Green's function

$$G_{[1,N]} = [A_N(\theta) - E]^{-1}. \tag{15.27}$$

We estimate $[A_N(\theta) - E]^{-1}(k_1, k_2)$ using the resolvent identity and (15.25). Precisely, if $0 \le j \le N$, then (15.25) holds for at least one of the matrices in the set

$$\{A_{n_0}(\theta+j\omega), A_{n_0-1}(\theta+j\omega), A_{n_0-1}\big(\theta+(j+1)\omega\big), A_{n_0-2}\big(\theta+(j+1)\omega\big)\}. \tag{15.28}$$

*Remark* 15.29. Observe that due to this last fact and endpoint considerations, we may have to replace in (15.27) $N$ by $N-1$ and (or) $\theta$ by $\theta+\omega$. We will come back to this point.

It is important what one obtains precisely for the off-diagonal decay of (15.27) by application of (15.25), (15.26) together with the resolvent identity. We will therefore go over the details.

Assume

$$L_{n_0} > 10^3\rho\log(1+\|v\|_\infty) \tag{15.30}$$

and denote by

$$\gamma = L_{n_0} - 70\rho\log(1+\|v\|_\infty) \tag{15.31}$$

the exponent in (15.26).

Taking (15.28), (15.29) into account, given $k_1, k_2 \in [1,N]$, we see that there is a size $n_0$-interval $\Lambda \subset [1,N]$ such that $A' = A_\Lambda$ satisfies (15.25) and $\{k \in [1,N] \mid |k-k_1| < \frac{n_0}{4}\} \subset \Lambda$. Thus from the resolvent identity and (15.25), (15.26), we get

$$|G_{[1,N]}(k_1,k_2)| < |G_\Lambda(k_1,k_2)|\chi_\Lambda(k_2) + \sum_{\substack{k_3\in\partial\Lambda, k_3'\in[1,N]\setminus\Lambda \\ |k_3-k_3'|=1}} |G_\Lambda(k_1,k_3)||G_{[1,N]}(k_3',k_2)| \tag{15.32}$$

$$< e^{-|k_1-k_2|L_{n_0}+7\rho n_0\log(1+\|v\|_\infty)}\chi_\Lambda(k_2) + \sum_{\substack{k_3\in\partial\Lambda, k_3'\in[1,N]\setminus\Lambda \\ |k_3-k_3'|=1}} e^{-\gamma|k_1-k_3|}\,|G_{[1,N]}(k_3',k_2)|. \tag{15.33}$$

We use here the fact that, by choice of $\Lambda$, $|k_1 - k_3| \geq \frac{n_0}{4}$ in the second term of (15.31).

From the preceding, it follows that

(15.34)
$$\max_{1 \leq k_1, k_2 \leq N} |G_{[1,N]}(k_1, k_2)| \leq e^{7\rho n_0 \log(1+\|v\|_\infty)} + e^{-\gamma \frac{n_0}{4}} \max_{1 \leq k_1, k_2 \leq N} |G_{[1,N]}(k_1, k_2)|.$$

Hence

$$\max_{k_1, k_2} |G_{[1,N]}(k_1, k_2)| < 2e^{7\rho n_0 \log(1+\|v\|_\infty)}. \tag{15.35}$$

Assume $|k_1 - k_2| > n_0$. The only contribution in (15.31) is the second term (15.33), bounded by

$$2e^{\gamma}\ e^{-\gamma|k_1 - k_3'|}\ |G_{[1,N]}(k_3', k_2)| \tag{15.36}$$

for some $k_3' \in [1, N]$, $\frac{n_0}{4} \leq |k_1 - k_3'| \leq n_0$. One easily verifies that iteration of (15.36) implies that if $|k_1 - k_2| > \frac{N}{10}$

$$\begin{aligned} |G_{[1,N]}(k_1, k_2)| &< (2e^{\gamma})^{10\frac{N}{n_0}}\ e^{8\rho n_0 \log(1+\|v\|_\infty)} e^{-\gamma|k_1 - k_2|} \\ &\overset{(15.30,31)}{<} e^{-\gamma(1 - \frac{n_0}{N} - \frac{200}{n_0})|k_1 - k_2|} \\ &< e^{-\gamma(1 - \frac{300}{n_0})|k_1 - k_2|}. \end{aligned} \tag{15.37}$$

In particular

$$\frac{1}{\|M_N(\theta)\|} \leq \frac{1}{|\det[A_N(\theta) - E|} = |G_{[1,N]}(1, N)| < e^{-\gamma(1 - \frac{300}{n_0})N}, \tag{15.38}$$

$$\frac{1}{N} \log \|M_N(\theta)\| > \left(1 - \frac{300}{n_0}\right)\gamma. \tag{15.39}$$

Coming back to Remark 15.29, observe that the same conclusion (15.39) holds if $N$ is replaced by $N - 1$ or $\theta$ by $\theta + \omega$. Thus (15.39) holds for all $\theta \notin S$. Recalling (15.17), (15.31) we obtain therefore

$$\begin{aligned} L_N &\geq \int_{\mathbb{T}^d \setminus S} \left[\frac{1}{N} \log \|M_N(\theta)\|\right] d\theta \\ &\geq (1 - e^{-n^{\sigma/5}})\left(1 - \frac{300}{n_0}\right)\big(L_{n_0} - 70\rho \log(1 + \|v\|_\infty)\big) \\ &> L_{n_0} - 71\rho \log(1 + \|v\|_\infty) \end{aligned} \tag{15.40}$$

by (15.4) and with $n$ large enough.

Recalling the construction of $n_0$ in Lemma 15.3, we get that

$$L_N > L_n - 71\rho \log(1 + \|v\|_\infty) \tag{15.41}$$

provided, cf. (15.30),

$$L_n > 10^3 \rho \log(1 + \|v\|_\infty), \tag{15.42}$$

and $v_0$ is a nonconstant real-analytic potential on $\mathbb{T}^d$.

*Step* 4 (Proof of Theorem 2). Let $v_0$ be a nonconstant real-analytic potential on $\mathbb{T}^d$.

LEMMA 15.43. *There is a constant* $c_0 = c_0(v_0) > 0$ *such that for small* $\delta$

$$\sup_{E_1 \in \mathbb{C}} \operatorname{mes}[\theta \in \mathbb{T}^d \big| \, |v_0(\theta) - E_1| < \delta] < \delta^{c_0}. \tag{15.44}$$

This may be deduced from Lojasiewicz' inequality. A proof also follows from an estimate in [B] based on the preparation theorem.

Next, we choose a large integer $n_1$. Then, we estimate for $\lambda > \lambda_0, \lambda_0$ sufficiently large,

$$\begin{aligned} \operatorname{mes}[\theta \in \mathbb{T}^d | \min_{1 \le j \le n_1} |v_0(\theta + j\omega) - \frac{E}{\lambda}| < \lambda^{-\frac{1}{100}}] \\ < n_1 \lambda_0^{-\frac{c_0}{100}} < \frac{1}{n_1}, \end{aligned} \tag{15.45}$$

invoking (15.44).

Since

$$M_{n_1}(\omega, \theta, E) = \lambda^{n_1} \prod_{n_1}^{1} \begin{pmatrix} v_0(\theta + j\omega) - \frac{E}{\lambda} & \frac{1}{\lambda} \\ -\frac{1}{\lambda} & 0 \end{pmatrix} \tag{15.46}$$

it follows from (15.45) that except for $\theta$ in a set of measure at most $\frac{1}{n_1}$,

$$\|M_{n_1}(\omega, \theta, E)\| > |\lambda|^{n_1} \left(|\lambda|^{-\frac{1}{100}} + 0(|\lambda|^{-1})\right)^{n_1} > |\lambda|^{\frac{98}{100} n_1}. \tag{15.47}$$

Thus

$$\log \lambda + 0(1) > L_{n_1} > \left(1 - \frac{1}{n_1}\right) \frac{98}{100} \log \lambda > \frac{97}{100} \log \lambda. \tag{15.48}$$

If we take $n = n_1$ in the preceding, the value of $\rho$ in (15.4) equals

$$\rho = \frac{\log \lambda + 0(1)}{L_{n_1}} (\log n_1)^{-1/2} \sim (\log n_1)^{-1/2} \tag{15.49}$$

and (15.42) holds in particular.

According to (15.14), let

$$n_2 = N = [e^{n_1^{\sigma/10}}] \tag{15.50}$$

for which, by (15.41), (15.49),

$$L_{n_2} > L_{n_1} - 71\rho\big(\log\lambda + O(1)\big) > L_{n_1} - 100\frac{\log\lambda}{(\log n_1)^{1/2}} \overset{(15.48)}{>} \left(1 - \frac{200}{(\log n_1)^{1/2}}\right)L_{n_1}. \tag{15.51}$$

The continuation of the process is clear. Assume

$$L_{n_j} > \frac{1}{2}\log\lambda. \tag{15.52}$$

Now, let

$$n = n_j,\ N = n_{j+1} = [e^{n_j^{\sigma/10}}]. \tag{15.53}$$

For $\rho$ in (15.4), we get $\rho \sim (\log n_j)^{-1/2}$ and (15.42) holds.

Thus (15.41) implies

$$L_{n_{j+1}} > L_{n_j} - \frac{1000}{(\log n_j)^{1/2}}\log\lambda > \left(1 - \frac{2000}{(\log n_j)^{1/2}}\right)L_{n_j}. \tag{15.54}$$

Iteration of (15.54) gives

$$L_{n_{j+1}} > \prod_{1\le j'\le j}\left(1 - \frac{2000}{(\log n_{j'})^{1/2}}\right)L_{n_1} > \left(1 - \frac{1}{100}\right)L_{n_1} \tag{15.55}$$

with (15.53) taken into account along with the fact that $n_1$ is large.

Since (15.48) holds, in particular,

$$L_{n_{j+1}} > \frac{99}{100}\cdot\frac{97}{100}\log\lambda > \frac{1}{2}\log\lambda. \tag{15.56}$$

Thus condition (15.52) follows. Also

$$L(\omega, E) = L = \inf_j L_{n_j} > \frac{1}{2}\log\lambda \tag{15.57}$$

proving Theorem 2.

*Remark.* An alternative proof of Theorem 2 appears in [G-S]. Their argument gives also an improved rate of convergence of $L_N(E)$ to $L(E)$ with consequences to the regularity of the integrated density of states $N(E)$.

Institute for Advanced Study, Princeton, NJ
*E-mail address*: bourgain@math.ias.edu

University of Toronto, Toronto, Ontario, Canada
*Current address*: Institute for Advanced Study, Princeton, NJ
*E-mail address*: mgold@math.ias.edu